\documentclass[a4paper,11pt]{article}
\pdfoutput=1 

\usepackage{jheppub} 

\usepackage[T1]{fontenc} 

\usepackage[utf8]{inputenc}
\usepackage{lmodern}
\usepackage{amsmath,amssymb}
\usepackage{graphicx}
\usepackage{url}
\usepackage{color}
\usepackage{enumitem}
\usepackage{subfigure}
\usepackage[dvipsnames]{xcolor}
\usepackage{bm}
\usepackage{epsfig}
\usepackage{amsmath}
\usepackage{amssymb}
\usepackage{slashed}
\usepackage{color}
\usepackage{accents}
\usepackage[dvipsnames]{xcolor}
\usepackage{tikz}
\usepackage{tabularx}

\newcommand{\exclude}[1]{}

\usepackage{braket}
\usepackage{url}

\usepackage{mathrsfs}

\newcommand{\tx}{\text}

\usepackage{fontawesome}

\usepackage{comment}

\newcommand{\bp}{\begin{pmatrix}}
\newcommand{\ep}{\end{pmatrix}}
\newcommand{\bb}{\begin{bmatrix}}
\newcommand{\eb}{\end{bmatrix}}

\newcommand{\df}{\text{d}}

\newcommand{\al}[1]{\begin{align}#1\end{align}}

\newcommand{\paren}[1]{\left(#1\right)}
\newcommand{\pn}[1]{\left(#1\right)}
\newcommand{\sqbr}[1]{\left[#1\right]}
\newcommand{\fn}[1]{\!\paren{#1}} 

\newcommand{\nn}{\nonumber\\}

\newcommand{\p}{\partial}

\usepackage{fancybox}

\usepackage{amsthm}
\theoremstyle{definition}

\newcommand{\ov}{\over}

\newcommand{\mc}{\mathcal}

\newcommand{\pr}{\prime}
\newcommand{\eV}{\text{eV}}

\newcommand{\MeV}{\text{MeV}}
\newcommand{\GeV}{\text{GeV}}

\newcommand{\CEnNS}{\text{CE}\nu\text{NS}}

\newcommand{\fm}{\text{fm}}

\newcommand{\g}{\text{g}}
\newcommand{\kg}{\text{kg}}

\usepackage{ulem} 

\title{Unfolding the low-energy reactor neutrino flux  from $\CEnNS$ data with a finite Dirac sum}

\author[a]{Muping Chen,}
\author[b]{Graciela Gelmini,}
\author[c]{Danny Marfatia,}
\author[b,1]{and Koichiro Yasuda\note{Corresponding author.}}

\affiliation[a]{International Center for Quantum-field Measurement Systems for Studies of the Universe and Particles (QUP, WPI), High Energy Accelerator Research Organization (KEK)\\ Oho 1-1, Tsukuba, Ibaraki
305-0801, Japan}
\affiliation[b]{Department of Physics and Astronomy, University of California, Los Angeles \\ Los Angeles, California, 90095-1547, USA}
\affiliation[c]{Department of Physics and Astronomy,
University of Hawaii at Manoa\\
Honolulu, HI 96822, USA}

\emailAdd{mpchen@post.kek.jp}
\emailAdd{gelmini@physics.ucla.edu}
\emailAdd{dmarf8@hawaii.edu}
\emailAdd{yasuda@physics.ucla.edu}

\abstract{We present a novel method to analyze coherent elastic neutrino-nucleus scattering ($\CEnNS$) data  to extract the reactor antineutrino spectrum below the inverse beta decay threshold of 1.8~MeV, where it remains unmeasured. Adapting halo-independent analysis techniques  developed for direct dark matter detection, we show how to obtain a best-fit and a pointwise confidence band for the integrated neutrino flux, without assuming a parametric form  or smoothness prior for the spectrum.  In our approach, which follows from convex geometry arguments, the differential neutrino rate is written as a linear combination of Dirac delta functions -- a finite Dirac sum (FDS) -- with a maximum number of terms determined by the number of data points.  We apply our ``FDS method'' to mock $\CEnNS$ data for a low-threshold Ge detector and compare it with Tikhonov-regularized unfolding.
}


\date{\today}

\begin{document}
\maketitle

\section{Introduction}

Inverse beta decay in scintillation detectors and Gadolinium-doped water Cherenkov detectors has been used to detect the electron antineutrinos emitted by  reactors, by taking advantage of the correlated signature of prompt positron emission followed by delayed neutron capture to significantly suppress backgrounds. However, this process has a threshold of 1.806~MeV.  
The sub-threshold reactor antineutrino spectrum remains unmeasured, and its theoretical prediction is subject to significant modeling uncertainties. Moreover, the neutron-capture component of the flux is still to be discovered. Individual models 
do not estimate the uncertainties in the flux, and are consistent with each other only at the 20\% level~\cite{Li:2001ha}.

 The observation of coherent elastic neutrino-nucleus scattering ($\CEnNS$) in recent years has opened a novel avenue for the study of low-energy neutrinos since this process has no kinematic 
 threshold~\cite{COHERENT:2017ipa, Colaresi:2022obx, NUCLEUS:2017htt, NUCLEUS:2019igx}.
Keeping in mind that approximately $70\%$ of the reactor antineutrino flux lies below $1.8~\MeV$, 
it has recently been proposed to use $\CEnNS$ to measure the reactor-antineutrino spectrum below the inverse beta decay threshold~\cite{Liao:2023kyy}.

The NUCLEUS experiment currently  aims to measure the $\CEnNS$ cross section using gram-scale cryogenic calorimeters with a nuclear-recoil energy threshold of ${\cal O}(10)$~eV~\cite{NUCLEUS:2026pnv}. 
A prototype has achieved a 20~eV threshold using a 0.5~g Al$_2$O$_3$ target~\cite{NUCLEUS:2017htt}.
Plans are underway for a technical run with a 7~g CaWO$_4$ target in the near future. Eventually, a 1~kg Ge detector, NUCLEUS-1kg, may achieve a threshold of 5~eV~\cite{NUCLEUS:2017htt}.

Using mock $\CEnNS$ scattering data for a NUCLEUS-like experiment~\cite{NUCLEUS:2019igx}, Ref.~\cite{Liao:2023kyy} applied Tikhonov-regularized unfolding to extract the low-energy reactor neutrino spectrum. This analysis assumed a flat background~\cite{NUCLEUS:2019igx} under the assumption that with the planned phased multi-target approach, the background will be well measured by the time NUCLEUS-1kg commences data-taking. The study of Ref.~\cite{Liao:2023kyy} concluded that the identification of the neutron capture component of the flux below about 1.8~MeV would require an unrealistically high exposure (of 300 kg$\cdot$year) even assuming a very low threshold (of 1~eV) and very small backgrounds (1 count per keV$\cdot$kg$\cdot$day).
 
In this work we investigate how a different neutrino spectrum extraction approach  compares with Tikhonov-regularized unfolding when analyzing the same mock scattering data as Ref.~\cite{Liao:2023kyy}, while allowing for more realistic backgrounds. Comparing our results with those of Ref.~\cite{Liao:2023kyy} for a particular choice of detector characteristics and background model, we find that to set an upper limit on the flux, as is essential to a determination of the neutron capture component, Tikhonov-regularized unfolding does better.  A lower limit on the flux below 0.8~MeV
is stronger with our method, and the pointwise confidence bands we obtain are comparable above 0.8~MeV. Also, with our analysis method, the identification of the neutron capture component of the flux below 1.8~MeV requires an unrealistically high exposure.

We call our method ``FDS'' because, based on  convex geometry arguments~\cite{Gelmini:2017aqe}, the differential neutrino spectrum is written as a finite Dirac sum, whose maximum number of terms is given by the number of data points.  As explained in Section~\ref{Sec:our-method}, our method was originally developed  for the   formally identical problem of determining the speed distribution of dark matter particles from direct detection data -- a framework known in that context as the Halo Independent method~\cite{Fox:2010bz,Fox:2011qd,Frandsen:2011gi,Gondolo:2012rs,Gelmini:2017aqe}.

The paper is organized as follows. In Section~\ref{Sec:our-method} we introduce our method and write the scattering rate as a convolution of a detector-dependent response function and the incoming neutrino flux. In Section~\ref{Sec:CEvNS-rate} we describe our detector modeling and how we generate our mock $\CEnNS$ data.
In Section~\ref{Sec:CGBprocedure}, we define and derive the response function, the essential kernel for our unfolding method. The separation of the low energy flux we are actually interested in reconstructing is done in Section~\ref{sec:low-energy-neutrino flux}. We explain our numerical procedure in Section~\ref{sec:numerical-procedure} and present our results in Section~\ref{Sec:Results}.
In Section~\ref{Sec:Discussion} we summarize our findings.

\section{The FDS method}
\label{Sec:our-method}  

Our method was originally developed to determine the local dark matter (DM) velocity distribution from direct detection data. To adapt it to our work, we start by establishing a formal analogy between the DM and neutrino applications.

In direct DM  detection experiments, local galactic dark halo particles interact within a detector causing nuclear recoils of energy $E_R$. If directionality is not important, as when measuring the time average event rate in nonpolarized targets, the particle flux depends on the DM speed distribution $F(v)$ (rather than the velocity distribution).  When considering nuclear recoils due to coherent neutrino interactions,  the neutrino flux $\phi(E_\nu)$ replaces $F(v)$ (and $E_\nu$ replaces $v$) in the calculation of the experimental event rate. This substitution forms the basis for the formal analogy between the two contexts.

While the usual ``halo-dependent'' (short for ``dark halo model dependent'') analysis of direct DM detection data requires assuming a model for $F(v)$ at the detector location to compute the expected event rate, the alternative and complementary ``halo-independent'' (HI) method instead uses the scattering data to infer the  DM speed (or velocity when needed) distribution (scaled by some factors, such as the local halo DM density).  The HI method was proposed using a simplified treatment~\cite{Fox:2010bz} and was soon extended~\cite{Gondolo:2012rs}
to fully incorporate experimental energy resolutions and efficiencies, nuclear form factors with arbitrary energy dependence, targets with any isotopic composition~\cite{DelNobile:2013cta}, and any type of DM interaction~\cite{DelNobile:2013cva}. 
In this more general approach~\cite{Gondolo:2012rs,DelNobile:2013cta,DelNobile:2013cva} the observed scattering  rate
is expressed as a convolution in the speed of a kernel -- which depends  only on the detector and the DM model -- and a (detector and particle model independent) function, proportional to $F(v)$ or its integral,  that contains all the information on the incoming particle distribution and is common to all experiments.  This ``halo-function''  is frequently chosen to be
$\eta(v_{\rm min})= \int_{v>v_{\rm min}}  dv {F(v)}/{v}$
(times some constants), where for a particular $E_R$,  $v_{\rm min}$ is the minimum speed of the DM particle that can produce it, which depends on the target mass.

Experiments do not directly measure $E_R$  but rather a proxy $E'$ for it (such as scintillation, heat or photoelectron count), related to $E_R$ by an energy resolution function. Thus, computing the predicted rate at a particular $E'$ involves  an integration in $E_R$. 
Since the relationship between $E_R$ and  $v_{\rm min}$ depends on the mass of the target nucleus, in targets with more than one nuclide (element or isotope), the mapping between $E_R$ and $v_{\rm min}$ is not unique, requiring the choice of one of them as an independent variable, which we choose to be $v_{\rm min}$~\cite{Gondolo:2012rs}. Then, a $v_{\rm min}$ value corresponds to different target nuclide dependent recoil energies, $E_R^N$, understood to be the maximum recoil energy for elastic scattering. This choice allows us to account for any isotopic target composition by summing terms dependent on $E_R^N (v_{\rm min})$ over target nuclides $N$, for any fixed detected energy $E'$, which are included in the detector ``response function'' (see below).

For DM scattering, we write the recoil rate as
\al{
    {\df R\ov \df E'} = \int_0^\infty 
    \df v~ {\p \mathcal{H} (v, E')\ov \p E'}
    \, {F(v)} = \int_0^\infty 
    \df v~ {\p \mathcal{R} (v, E')\ov \p E'}
    \, {\eta(v)}\,.
\label{eq:HI-1}
}

The same arguments apply to neutrino scattering by replacing the DM speed and distribution by the incoming neutrino energy $E_\nu$ and flux $\phi(E_\nu)$, so that the differential rate becomes\footnote{A strict analogy would require $\phi(E_\nu)$ to be normalized to unity as is $F(v)$. The normalization factor can instead be applied to the response function without changing any of our results.}
\al{
    {\df R\ov \df E'} = \int_0^\infty 
    \df E_\nu~ {\p \mathcal{H}(E_\nu, E')\ov \p E'}{\phi(E_\nu)} 
    = \int_0^\infty 
    \df E_\nu~ {\p \mathcal{R}(E_\nu, E')\ov \p E'} {\Phi(E_\nu)} \,,
\label{eq:FI-1}
}
where $\Phi\fn{E_{\nu}}$ is the  neutrino flux integrated above $E_\nu$:
\al{
    \Phi\fn{E_{\nu}} &= \int_{E_\nu}^\infty\df \tilde{E}_\nu~ \phi\fn{\tilde{E}_{\nu}}\,.
    \label{integrated-nu-flux}
}
In Eqs.~\eqref{eq:HI-1} 
and~\eqref{eq:FI-1} the right-most expressions are derived from those to their left by integration by parts, and $\p \mathcal{H}/\p E'$ and $\p \mathcal{R}/ \p E'$ (where $\mathcal{R}=\partial \mathcal{H}/\partial v$ in Eq.~\ref{eq:HI-1},  and  
 $\mathcal{R}=\partial \mathcal{H}/\partial E_\nu$ in Eq.~\ref{eq:FI-1}) are kernels that depend strictly on the particle interaction model and detector characteristics. We will refer to them as detector ``response functions''.  Notice that they act as ``window functions'' in speed or neutrino energy, respectively,  since  the functions of interest  can be measured by an observed rate at $E'$ only within the range of $v$ or $E_\nu$ in which the kernels are nonzero. This range is determined by the energy resolution function and the relation between the recoil energy $E_R$ and $v$ or $E_\nu$, respectively.  

Having established the formal analogy between the DM and neutrino applications that allows us to adapt the HI method to neutrino scattering, we focus on neutrinos in the following. 

Any likelihood
 depends on the incoming differential flux ${\phi(E_\nu)}$, or integrated flux ${\Phi(E_\nu)}$,   
 through the expected rates, which, as we showed, are convolutions over these functions.    While conventional likelihood methods are designed for parameter estimation, our goal here is to extract an entire function without assuming its form.
 
In Refs.~\cite{Fox:2014kua, Gelmini:2015voa}, for an extended likelihood function for unbinned data, it was proven (using Karush-Kuhn-Tucker conditions, extended to functionals~\cite{Gelmini:2015voa}) that the best-fit ${\Phi(E_\nu)}$ was  a piecewise constant function with a number of downward steps not larger than the number of data points. In addition, a two-sided pointwise confidence band was rigorously defined at any desired confidence level (CL).  The fact that the best fit was such a peculiar function was initially puzzling, and the analysis could not be extended to other types of likelihood functions until concepts of convex geometry were applied~\cite{Gelmini:2017aqe}. By applying the Fenchel-Eggleston theorem (an extension of the
Caratheodory theorem) to the convex hull of even rates, whose generating vectors are the response functions,
Ref.~\cite{Gelmini:2017aqe} showed that given a number $d$ of data points, any likelihood can be maximized with (a speed distribution for DM, or here) a differential neutrino flux that is a finite Dirac sum,\footnote{In cases where directionality is important, i.e., where the DM  velocity distribution in the HI method, or the neutrino flux as a function of the neutrino momentum vector $\vec{p}_\nu$ in the present application, is relevant, the sum in Eq.~\eqref{eq:FI-sum-deltas} is over delta functions in three dimensions,  ${\phi(\vec{p}_\nu)} = \sum_{s=1}^{d-1} \phi_s~ \delta(\vec{p}_\nu-\vec{p}_{\nu s})$, which has $4(d-1)$ parameters.}  
   \al{
    {\phi(E_\nu)} = \sum_{s=1}^{d-1} \phi_s~ \delta(E_\nu-E_{\nu s})\,,
\label{eq:FI-sum-deltas}
}
  where $\phi_s$ and $E_{\nu s}$ are 2$(d-1)$ constants to be determined by fitting the data. The data points could consist of single scattering events for unbinned data, in which case $d$ is the number of single events,  or counts per bin for binned data, in which case $d$ is the number of bins.  The integral of each Dirac delta in this linear combination, as a function of its lower limit of integration,  is a left-Heaviside function $H(E_{\nu s}-E_\nu)$. As a result, the integrated flux, 
   \al{
    {\Phi(E_\nu)} = \sum_{s=1}^{d-1} \phi_s~ H(E_{\nu s}-E_\nu) \,,
\label{eq:FI-sum-deltas}
}
is a piecewise-constant function with at most $d-1$ downward steps.

To work with a function whose value is defined at most energies, we fit the data with the integrated flux $\Phi(E_\nu)$.
   Since any likelihood can be maximized with $\Phi(E_\nu)$ written as in Eq.~\eqref{eq:FI-sum-deltas}, the problem of finding the function $\Phi(E_\nu)$ becomes one of finding the at most 2$(d-1)$ parameters $\phi_s$ and $E_{\nu s}$.
Although any likelihood can be maximized with the function in Eq.~\eqref{eq:FI-sum-deltas},  there may be other functions that maximize it as well, in which case there is a ``degeneracy band'' around the best-fit, as explained in Ref.~\cite{Gelmini:2017aqe}.
This is what we find in the present application.  

We use mock binned data with much larger than 30 events in each bin, so that a Gaussian likelihood function $\mathcal{L}$ is a good approximation to a Poisson likelihood. 
Minimizing  $ L = - 2 {\rm ln} \mathcal{L}$ for a Gaussian distribution is equivalent to minimizing the Neyman chi-squared statistic, 
\al{
    \chi^2_{\tx{Neyman}} &= 
    \sum_{i=1}^{d}{\pn{N_{\tx{exp-i}} - N_{\tx{obs-i}}}^2\ov N_{\tx{obs-i}}}\,,
}
since $L$ and $\chi^2_{\tx{Neyman}}$ differ by a constant.
This choice of statistic leads to faster numerical calculations than the other options we tested. 
Here $N_{\tx{exp-i}}$ and $N_{\tx{obs-i}}$ are respectively, the expected and observed number of $\CEnNS$ events in each bin $i$. The definition of 
$\chi^2_{\tx{Neyman}}$ specific to our application is given in Eq.~\eqref{chiSq}.

To find the best-fit function of the form in Eq.~\eqref{eq:FI-sum-deltas} we adopt a procedure proposed 
in Ref.~\cite{Feldstein:2014gza}. 
As in our current procedure, Ref.~\cite{Feldstein:2014gza} used an expansion in Dirac delta functions, although without knowing the maximum possible number of terms in the expansion. This was determined in Ref.~\cite{Gelmini:2017aqe}.
 We find a pointwise confidence band at each energy $E_\nu$  with the constrained minimization procedure in Section~6 of Ref.~\cite{Gelmini:2017aqe}, in which the integrated rate at a particular energy is fixed at a value 
above or below the best fit. 
In our study, the profile likelihood ratio is $\Delta L_{\rm min}= \Delta \chi^2_{\rm min}$.  
Because we find a degeneracy band, Wilks' theorem does not apply, and we use Monte Carlo methods to find the confidence level associated with the values of $\Delta \chi^2_{\rm min}$. 

For our application to CE$\nu$NS, we unfold
an integrated flux $\delta\Phi\fn{E_\nu}$ that differs from $\Phi\fn{E_\nu}$ by a constant. 
The reasons for this will become clear in Section~\ref{sec:low-energy-neutrino flux}. The method proceeds as follows. Divide the relevant neutrino energy range $E_\nu$ into a large number $N_{\rm int}$ of equally spaced intervals $j=1 \dots  N_{\rm int}$, in each of which $\delta\Phi(E_\nu)$ takes a constant value, $\delta\Phi_j$, as shown in Fig.~\ref{neutrino-flux-hist}. With this ansatz for $\delta\Phi(E_\nu)$, $\chi^2_{\tx{Neyman}}$ becomes a function of the $N_{\rm int}$ parameters $\delta\Phi_j$. The best-fit $\delta\Phi_j$ values, $\delta\Phi^{\rm Best\,fit}_j$, are determined by minimizing $\chi^2_{\tx{Neyman}}$ while imposing the physical requirement that $\delta\Phi(E_\nu)$ be nonincreasing, i.e., that the $\delta\Phi_j$ values cannot increase with increasing $j$. In the best-fit function, many adjacent intervals have identical $\delta\Phi^{\rm Best\,fit}_j$ values, forming a constant segment, resulting in no more than $d-1$ downward steps. 

To choose the number of intervals, starting from a particular $N_{\rm int}$ value for our initial fit, 
we increase (e.g. double) it and find the new best fit.  If the initial $N_{\rm int}$ value is not much larger than the number of data points, increasing $N_{\rm int}$ leads to a rapid decrease of $\chi^2_{\rm min}$ (as in Fig.~\ref{chiSquaredComparison}). We continue increasing $N_{\rm int}$
 until a further increase of this number ceases to produce a significant decrease in $\chi^2_{\rm min}$ or a change in the shape of the best-fit function. It is necessary to take a large enough number of intervals $N_{\rm int}$ to determine the precise locations of the steps (which must be at the boundaries of the intervals), and after finding these locations, doubling $N_{\rm int}$ does not move them.\footnote{Reference~\cite{Feldstein:2014gza} pointed out that the number of downward steps found in this manner always remains smaller than the number of data bins, a result that was later explained in  Ref.~\cite{Gelmini:2017aqe}.} 

Note that the procedure to find the picewise constant best-fit function is not unique. Our procedure is different from, for example, the global extremization algorithm used in Ref.~\cite{Gelmini:2015voa}.\footnote{The algorithm, in which the logarithms of the heights of the piecewise constant pieces  are fitted together with the location of the downward steps, was implemented in the CoddsDM publicly available software~\cite{Georgescu:CoddsDM}.}

\section{Detector modeling and mock $\CEnNS$ data}
\label{Sec:CEvNS-rate}

The differential recoil rate per unit detector mass for a nuclear recoil energy $E_R$, summed over nuclides $N$ in the target, is
\al{
    {\df R\ov \df E_R}\fn{E_R} = \sum_\tx{N} {\df R_{\rm N} \ov \df E_R }= \sum_\tx{N} {f_{\rm N}\ov M_{\rm N}}\int \df E_\nu~\phi\fn{E_\nu}\,{\p \sigma_{\rm N}\ov \p E_\text{R}}\fn{E_R, E_\nu}\,,
\label{eq:dR-dER}
}
where $\df R_{\rm N}/\df E_R$ is the per-nuclide contribution, $M_{\rm N}$ is the nuclide mass,  $f_{\rm N} = M^{\rm N}_T/M_T$ is the mass fraction of nuclide $N$ in the target of total mass $M_T$ (so that $f_{\rm N}/M_{\rm N}$ counts the number of  nuclides $N$ per unit target mass), and $\partial \sigma_{\rm N}/\partial E_{\rm R}$ is the cross section in Eq.~\eqref{crosssection}.

As explained in Appendix~\ref{AppA:cross-section}, both for simplicity and to compare our results with those of Ref.~\cite{Liao:2023kyy}, we model the target as pure $^{72}$Ge, whose mass number provides a good approximation to the standard atomic weight of natural Ge, 72.63~\cite{IUPAC}. With this choice
only one term contributes to the summation over $N$ with $f_{\rm N} = 1$ and $M^{\rm N}_T= M_T$. 

\begin{figure} 
    \centering
    \includegraphics[width=0.7\linewidth]{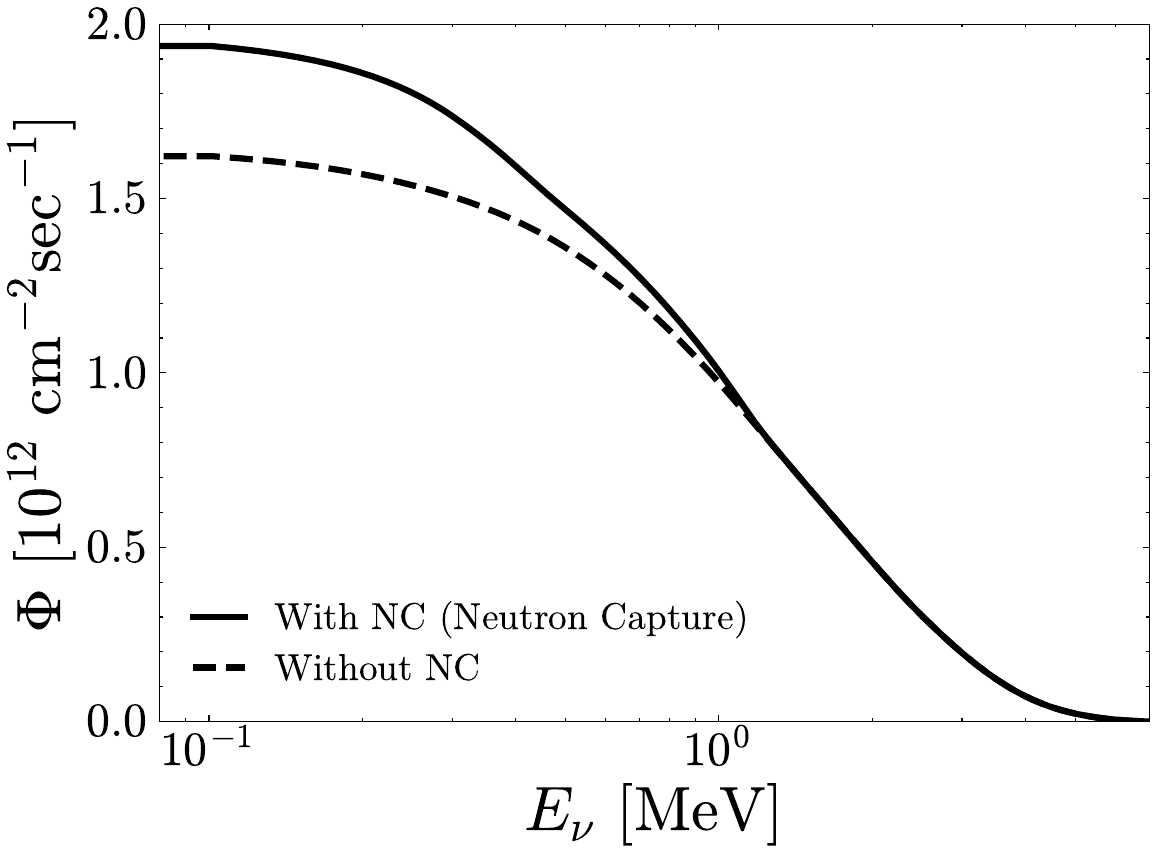}
    \caption{Predictions for the integrated reactor antineutrino flux $\Phi(E_\nu)$ with (solid curve) and without (dashed curve) the $^{238}$U neutron-capture component (see Appendix~\ref{AppB:theoretical-nu-flux} for details).}
    \label{fig-Int-Flux}
\end{figure}

The differential antineutrino flux $\phi(E_\nu)$ is theoretically modeled in Appendix~\ref{AppB:theoretical-nu-flux}.  Two theoretical  predictions for the integrated neutrino flux $\Phi$ above $E_\nu$, defined in Eq.~\eqref{integrated-nu-flux}, 
 are shown in Fig.~\ref{fig-Int-Flux}. The solid curve shows a prediction that includes the $^{238}$U neutron-capture (NC) component, whose discovery is a major goal. 
 This is the neutrino flux used to compute our mock  event rate data.
 We will determine what exposure for a given threshold is required to distinguish the solid curve from the dashed curve in Fig.~\ref{fig-Int-Flux}, which shows the prediction without the $^{238}$U NC component.

Experiments do not measure the recoil energy $E_R$ directly, but its proxy $E^\pr$ (a phonon signal in NUCLEUS), related to $E_R$ by the experimental energy resolution function, denoted by $G_{\rm N}(E^\pr, E_R)$. We are therefore interested in the predicted differential  event rate as a function of $E^\pr$, 
\al{
    {\df R\ov \df E^\pr}\fn{E^\pr} = \epsilon(E^\pr) \sum_{\rm N} \int_0^\infty \df E_R\, G_{\rm N}\fn{E^\pr, E_R}\, {\df R_{\rm N}\ov \df E_R}\fn{E_R}\,,
\label{eq:dR-dEprime}
}
where $\epsilon(E^\pr)$ is the detector efficiency, which we take to be a step function, $\epsilon(E^\pr) = \theta(E^\pr - E^\pr_{\tx{thr}})$.  For the cryogenic Ge detector considered here, we assume a Gaussian resolution function with width $\sigma_E = 1~\eV$,
\al{
    G_{\rm N}\fn{E^\pr, E_R} = {1\ov \sqrt{2\pi}\,\sigma_E}\,\exp\!\sqbr{-{(E^\pr - E_R)^2 \ov 2\sigma_E^2}}\,.
\label{eq:G-gauss}
}
 We checked that replacing this resolution function by a box of comparable width or by a wider Gaussian, and replacing the efficiency function by one with a smoother turn-on, does not significantly change our results.

The width $\sigma_E = 1~\eV$ is consistent with the resolution achievable by phonon-mediated cryogenic detectors. The NUCLEUS collaboration has reported a baseline phonon-energy resolution of $2.94 \pm 0.05~\eV$ in a CaWO$_4$ prototype with sensitivity-enhancement techniques applied~\cite{NUCLEUS:2024sensitivity}, and consistently below $10~\eV$ across multiple cool-downs~\cite{NUCLEUS:2017htt, NUCLEUS:2019igx}. 

We model the detector configuration after the NUCLEUS experiment~\cite{NUCLEUS:2017htt, NUCLEUS:2019igx}, with a cryogenic germanium target of mass $M_T=1$~kg. We consider two values of the measured energy threshold, $E^{\pr}_{\tx{thr}} = 1~\eV$ and $5~\eV$, in the range relevant for next-generation cryogenic detectors. The $5~\eV$ choice corresponds to a target for cryogenic Ge bolometers~\cite{NUCLEUS:2017htt}. The Ricochet collaboration in a Ge bolometer aims at a sub-100~eV nuclear-recoil reach in its CryoCube array deployed at the ILL reactor~\cite{Ricochet:2024CryoCube}.
NUCLEUS ${\rm Al_2O_3}$ calorimeters have already reached a threshold for nuclear recoils below $20~\eV$ in prototype operation~\cite{NUCLEUS:2017htt}, so a few-eV threshold is an optimistic extrapolation for a Ge target with comparable phonon-sensor performance. As we explain later, a 5~eV threshold reaches down to a neutrino energy of 0.4~MeV, leaving a large portion of the neutron-capture component undetectable. This is the main reason to consider a lower detector threshold of 1~eV.
The exposure $\mc E$ (target mass times running time, in
kg$\cdot$year) and the external background noise model are varied in the experimental scenarios we consider.

\begin{figure}
    \centering
    \includegraphics[width=0.7\linewidth]{
    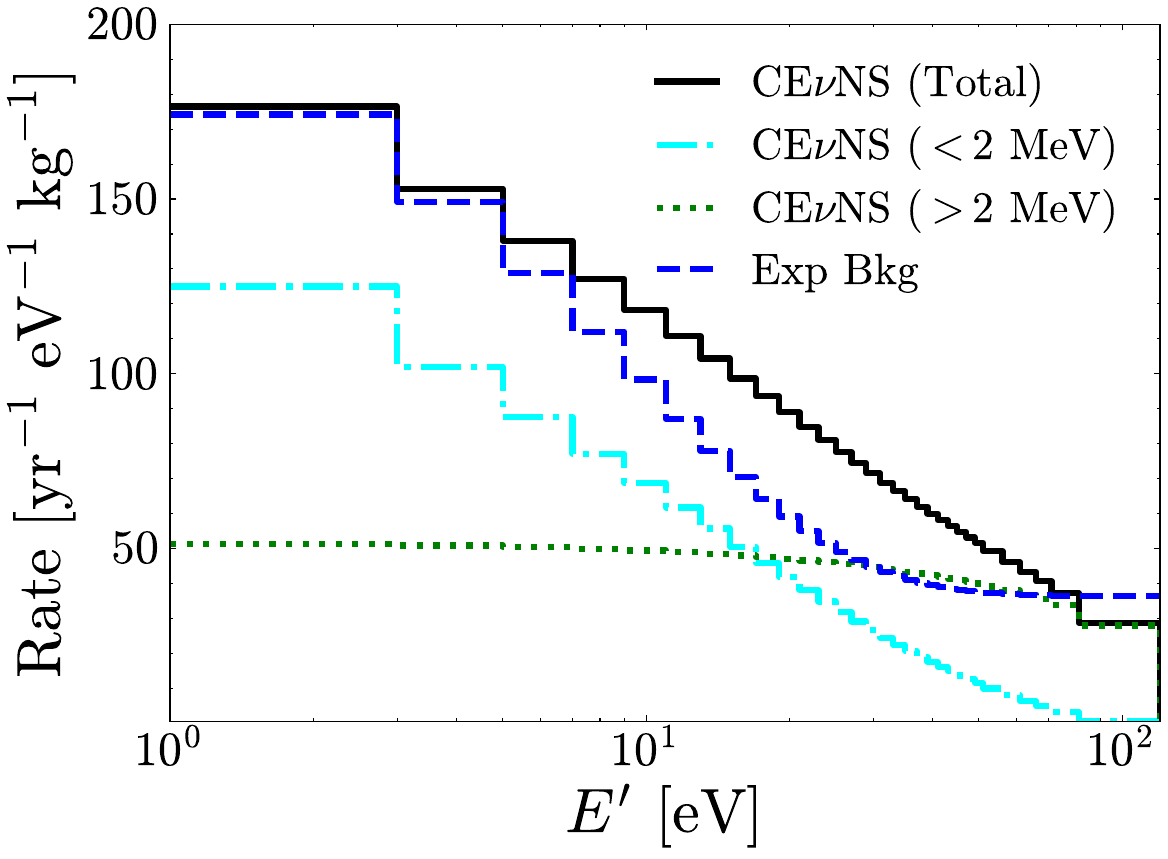}
    \caption{
    Binned event rates per unit detector mass as a function of the observed energy $E'$.  Four rates are shown: 1) the total $\CEnNS$ rate (black solid line), computed using Eq.~\eqref{eq:dR-dEprime} with the neutrino flux indicated by the solid line in Fig.~\ref{fig-Int-Flux}, 2) the  
    $\CEnNS$ contribution from neutrinos with $E_\nu < 2~\MeV$ (cyan dash-dotted line),  
    3) the $\CEnNS$ contribution from $E_\nu > 2~\MeV$ (green dotted line), and 4) the exponential background, ``Exp Bkg'', of Eqs.~\eqref{bkg-form} and~\eqref{bkg} (blue dashed line). 
     }
    \label{EventRate}
\end{figure}

With the $\phi(E_\nu)$ modeled as in Appendix~\ref{AppB:theoretical-nu-flux} and Eqs.~\eqref{eq:dR-dER} and~\eqref{eq:G-gauss}, the differential rate $\df R/\df E^\pr$ obtained from Eq.~\eqref{eq:dR-dEprime} is
shown in Fig.~\ref{EventRate}.  Here we use the same $E'$ data binning of Ref.~\cite{Liao:2023kyy}, with 25 bins of 2~eV width below 51~eV, 4~bins of 5~eV width between 51~eV and 71~eV, one bin from 71~eV to 81~eV, and one bin from 81~eV to 120~eV.

The signal from which we extract the low-energy neutrino spectrum consists of $\CEnNS$ events shown by the 
dash-dotted cyan line. We denote the $\CEnNS$ contribution in each bin due to neutrinos with $E_\nu > 2~\MeV$ by $h_i$, and display it with a dotted green line in Fig.~\ref{EventRate}. It is well constrained by inverse beta decay measurements.  The black line corresponds to the total number of $\CEnNS$ events in each bin.
A large exponential background defined in Eqs.~\eqref{bkg-form} and~\eqref{bkg} is shown as a dashed blue line. 

\section{Response functions }
\label{Sec:CGBprocedure}

We now define the detector-dependent kernel or response function needed for the convolution in Eq.~\eqref{eq:FI-1}. 
Substituting Eq.~\eqref{eq:dR-dER} for $\df R_{\rm N}/\df E_R$,  into Eq.~\eqref{eq:dR-dEprime} gives the measured rate as a double integral over $E_R$ and $E_\nu$. The kinematics of $\CEnNS$ determines $E_R$ as a function of $E_\nu$ for each nuclide, and the order of integration can be exchanged, taking into account the kinematic limits for elastic scattering. Thus, taking $E_{\rm R}^{\rm min}=E^\pr_{\rm thr}$,
\al{
    {\df R\ov \df E^\pr}&\fn{E^\pr} = \epsilon\fn{E^\pr}\sum_\tx{N} \int_0^\infty \df E_\tx{R}\, G_\tx{N}\fn{E^\pr, E_\tx{R}}\sqbr{{f_{\rm N}\ov M_{\rm N}}\int_{E_\nu > E_{\nu,{\rm N}}^{\rm min}(E_R)} \df E_\nu\, \phi\fn{E_\nu}\, {\p \sigma_{\rm N}\ov \p E_\tx{R}}\fn{E_\tx{R}, E_\nu}}\nn
    &= \int_0^\infty \df E_\nu~\phi\fn{E_\nu}\sum_\tx{N}\sqbr{{f_{\rm N}\ov M_{\rm N}}\,\epsilon\fn{E^\pr}\int_{E_{\tx{R}}^{\tx{min}}}^{E_{{\rm R, N}}^{\rm max}(E_\nu)} \df E_\tx{R}\, G_\tx{N}\fn{E^\pr, E_\tx{R}}\, {\p \sigma_{\rm N}\ov \p E_\tx{R}}\fn{E_\tx{R}, E_\nu}}\,.
\label{recoilRate1}
} 
The minimum neutrino energy required to produce a recoil $E_R$ on nuclide $N$ is
\al{
 E_{\nu, {\rm N}}^\tx{min}\fn{E_\tx{R}} = {{E_\tx{R} + \sqrt{E_\tx{R}\pn{E_\tx{R} + 2 M_\tx{N}}}}\ov 2}\,,
}
and conversely, the maximum recoil energy that a neutrino of energy $E_\nu$ can deposit on nuclide $N$ is
\al{
    E_{\tx{R}, {\rm N}}^\tx{max}\fn{E_\nu} = {2 E_\nu^2\ov M_\tx{N} + 2 E_\nu}\,.
\label{maxRecoil}
}
The square bracket in the second line of Eq.~\eqref{recoilRate1} is the per-nuclide response function for the differential flux,
\al{
    {\p\mc H_{\rm N}\ov\p E^\pr}\fn{E_\nu, E^\pr} = {f_{\rm N} \ov M_{\rm N}}\,\epsilon\fn{E^\pr}\int_{E_{\tx{R}}^{\tx{min}}}^{E_{\tx{R}, {\rm N}}^\tx{max}\fn{E_\nu}} \df E_\tx{R}\, G_\tx{N}\fn{E^\pr, E_\tx{R}}\, {\p \sigma_{\rm N}\ov \p E_\tx{R}}\fn{E_\tx{R}, E_\nu}\,,
    \label{partialdcurlyH} 
}
and the total response function is the sum over nuclides,
\al{
    {\p\mc H\ov\p E^\pr}\fn{E_\nu, E^\pr} = \sum_{\rm N} {\p\mc H_{\rm N}\ov\p E^\pr}\fn{E_\nu, E^\pr}\,.
    \label{eq:dcurlyHdE'}
}
This kernel acts as a window in $E_\nu$ through which the measured differential rate can give information on the neutrino flux. It is nonvanishing only over the range of incoming neutrino energies that contributes to recoils within the experimental resolution at the given measured energy $E^\pr$. With the definition in Eq.~\eqref{eq:dcurlyHdE'}, the differential event rate in Eq.~\eqref{recoilRate1} can be written in the form of Eq.~\eqref{eq:FI-1}.
\al{
    {\df R\ov \df E^\pr}\fn{E^\pr} = \int_{0}^{\infty}\df E_\nu\,\phi\fn{E_\nu}\,{\p\mc H\ov\p E^\pr}\fn{E_\nu, E^\pr}\,.
\label{eq:dR-dEprime-H}
}

The right-most expression in Eq.~\eqref{eq:FI-1} is derived integrating Eq.~\eqref{eq:dR-dEprime-H}  by parts in $E_\nu$, 
\al{
    {\df R\ov \df E^\pr}\fn{E^\pr} = -\sqbr{\Phi\fn{E_{\nu}}\,{\p\mc H\ov\p E^\pr}\fn{E_{\nu}, E^\pr}}_0^\infty + \int_0^\infty\df E_{\nu}\,\Phi\fn{E_{\nu}}\,{\p\ov\p E_\nu}\sqbr{{\p\mc H\ov\p E^\pr}\fn{E_\nu, E^\pr}}\, ,
    \label{IntByParts}
}
where $\Phi(E_\nu)$ is the  integrated flux  defined in Eq.~\eqref{integrated-nu-flux}.
The boundary term in Eq.~\eqref{IntByParts} vanishes because there are no incident neutrinos with $E_\nu \to \infty$, so $\Phi \to 0$, while as $E_\nu \to 0$ the kinematic upper limit $E_{\rm R}^{\max}(E_\nu) \to 0$ collapses the integration domain in the response function so that $\partial\mc H/\partial E^\pr \to 0$. Defining the response function that acts on the integrated flux as
\al{
    {\p\mc R\ov \p E^\pr}\fn{E_\nu, E^\pr} \equiv {\p\ov \p E_\nu}\sqbr{{\p\mc H\ov \p E^\pr}\fn{E_{\nu}, E^\pr}}\,,
\label{eq:R-kernel}
}
we obtain,
\al{
    {\df R\ov \df E^\pr}\fn{E^\pr} = \int_0^\infty\df E_{\nu}\,\Phi\fn{E_{\nu}}\,{\p\mc R\ov \p E^\pr}\fn{E_\nu, E^\pr}\,.
\label{eq:integral-transform}
}
We will use this equation to reconstruct
 the antineutrino spectrum from data. Since the mock data shown in Fig.~\ref{EventRate} are binned, the rate over each bin must be integrated to obtain the number of events per bin.

\begin{figure*}
    \centering
    \includegraphics[width=0.7\linewidth]{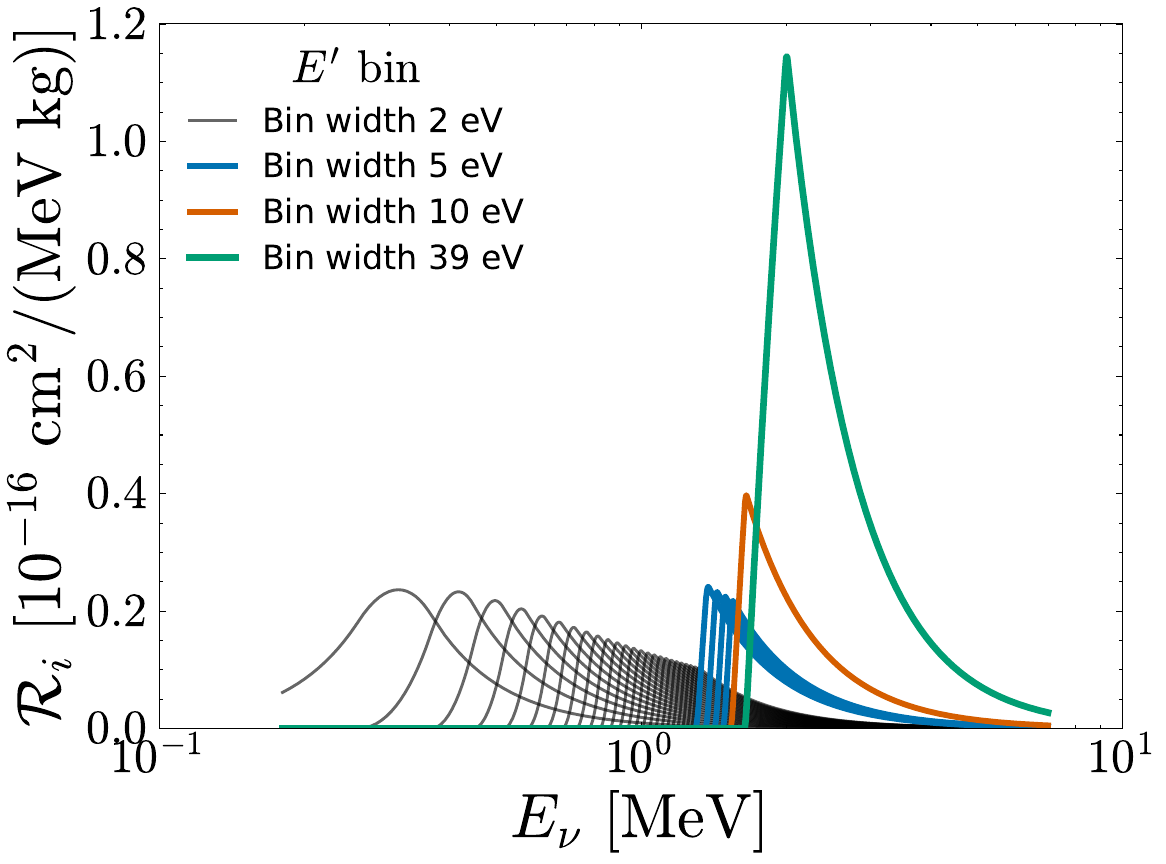}
    \caption{Integrated response functions $\mc R_i =\mc R_{\sqbr{E^\pr_i, E^\pr_{i+1}}}\fn{E_\nu}$ defined in Eq.~\eqref{Bin-IntegratedCurlyR} for $E^\pr_{\rm th}= 1$~eV. The black curves show the 25 bins from 1~eV to 51~eV, each of width 2~eV; the colored curves show the six wider bins, 51--56~eV, 56--61~eV, 61--66~eV, 66--71~eV, 71--81~eV, and 81--120~eV. For $E^\pr_{\rm th}= 5$~eV the response functions are identical except that the 1--3~eV and 3--5~eV bins fall below threshold and are absent.  
    }
    \label{fig:IntegratedCurlyR}
\end{figure*}

The $E^{\pr}$ range of interest is from the experimental threshold $E^{\pr}_{\tx{thr}}$ to 120~eV, the maximum measurable recoil energy for $E_\nu \leq 1.8$~MeV (given by Eq.~\eqref{maxRecoil}).
We divide this range into $d$ bins that we label by the lowest energy in each bin $E^{\pr}_i$, $i = 1, \ldots, d$.
Following Ref.~\cite{Liao:2023kyy} we use $d=31$ bins
for $E^{\pr}_{\tx{thr}}=$ 1 eV, and two bins less for $E^{\pr}_{\tx{thr}}=$ 5 eV, as explained in the previous section. 

The predicted number of events in the $i^{\rm th}$ bin is
\al{
    N_i = \mc E \int_{E^{\pr}_i}^{E^{\pr}_{i+1}} \df E^\pr\, {\df R\ov \df E^\pr}\fn{E^\pr} = \mc E R_{\sqbr{E^\pr_i, E^\pr_{i+1}}}\,,
\label{recoilrateSim}
}
where    
\al{
    R_i \equiv R_{\sqbr{E^\pr_i, E^\pr_{i+1}}} \equiv \int_0^\infty \df E_\nu\, \Phi\fn{E_\nu}\, \mc R_{\sqbr{E^\pr_i, E^\pr_{i+1}}}\fn{E_\nu}\,,
\label{masterEquation}
}
 in terms of the bin-integrated response function,
\al{
    \mc R_i(E_\nu) \equiv \mc R_{\sqbr{E^\pr_i, E^\pr_{i+1}}}\fn{E_\nu} = \int_{E^\pr_i}^{E^\pr_{i+1}} \df E^\pr\, {\p\mc R\ov \p E^\pr}\fn{E_\nu, E^\pr}\,.
    \label{Bin-IntegratedCurlyR}   
}
These functions are shown in Fig.~\ref{fig:IntegratedCurlyR} for the mock data in Fig.~\ref{EventRate} with a 1~eV threshold.

\section{CE$\nu$NS rate from low energy neutrinos}
\label{sec:low-energy-neutrino flux}

 The formulation just presented -- which strictly reflects the earlier application to the dark matter context -- needs to be slightly modified because we are interested in reconstructing the  reactor antineutrino flux only below the inverse-beta decay threshold, rather than over all energies for which the bin-integrated response functions in Eq.~\eqref{Bin-IntegratedCurlyR} and Fig.~\ref{fig:IntegratedCurlyR} are significantly nonzero; this alternative is the subject of Appendix~\ref{AppD:Method-1}.  As in Ref.~\cite{Liao:2023kyy},
the upper end of the energy range of interest is 2~MeV instead of 1.8~MeV due to the way theoretical spectra are presented (see Appendix~\ref{AppB:theoretical-nu-flux}), and the reactor neutrino spectrum above 2~MeV is treated as a known measured input (cf.\ the high-energy contribution $h_i$ in Eq.~\eqref{eq:Ni-predicted}).
The contribution of high energy neutrinos with $E_\nu>2$~MeV to the rate in Eq.~\eqref{eq:dR-dEprime-H} at 2~MeV is
\begin{equation} \label{eq:rate-separated-at 2MeV}
    \frac{dR}{dE'}(E')=\int_0^{2\,{\rm MeV}} dE_{\nu}\phi(E_{\nu})\frac{\partial \mathcal{H}}{\partial E'}(E_{\nu},E')+\int_{2\,{\rm MeV}}^\infty dE_{\nu}\phi(E_{\nu})\frac{\partial \mathcal{H}}{\partial E'}(E_{\nu},E')\,,
\end{equation}
where the second term, 
\al{ \label{eq:h-definition}
    h\fn{E^\pr} &\equiv \int_{2\,{\rm MeV}}^\infty dE_{\nu}\phi(E_{\nu})\frac{\partial \mathcal{H}}{\partial E'}(E_{\nu},E')\,,
}
is known.
The first term contains the function we want to reconstruct from data.
Integrating this term by parts, we obtain
\al{ \label{int-by-parts-2MeV}
    &\int_0^{2\,{\rm MeV}} dE_{\nu}\phi(E_{\nu})\frac{\partial \mathcal{H}}{\partial E'}(E_{\nu},E')\nn
    &=-\left[ \Phi(E_{\nu})\frac{\partial \mathcal{H}}{\partial E'}(E_{\nu},E') \right]_0^{2\,{\rm MeV}}+\int_0^{2\,{\rm MeV}}dE_{\nu}\Phi(E_{\nu})\frac{\partial \mathcal{R}}{\partial E'}(E_{\nu},E') \\
    &=-\Phi(2\,{\rm MeV})\int_0^{2\,{\rm MeV}}dE_{\nu}\frac{\partial \mathcal{R}}{\partial E'}(E_{\nu},E')+\int_0^{2\,{\rm MeV}}dE_{\nu}\Phi(E_{\nu})\frac{\partial \mathcal{R}}{\partial E'}(E_{\nu},E')\notag\\
    &=\int_0^{2\,{\rm MeV}}dE_{\nu}[\Phi(E_{\nu})-\Phi(2\,{\rm MeV})]\frac{\partial \mathcal{R}}{\partial E'}(E_{\nu},E')\,,\notag
}
where, in the boundary term in the second line we used
\al{
\frac{\partial\mathcal{H}}{\partial E'}(2\,\mathrm{MeV},E') = \int_0^{2\,\mathrm{MeV}} \df E_\nu\, \frac{\partial \mathcal{R}}{\partial E'}(E_{\nu},E')\,, 
}
which follows from Eq.~\eqref{eq:R-kernel} and the vanishing of $\partial\mathcal{H}/\partial E'$ at $E_\nu = 0$.
Unlike the boundary term in Eq.~\eqref{IntByParts}, which vanishes at both limits, here the upper limit is nonzero: $\Phi(2\,\mathrm{MeV})= 0.458 \times 10^{12}/($cm$^2$s). Defining
\begin{equation} 
  \delta\Phi\fn{E_\nu} \equiv \Phi\fn{E_\nu} - \Phi\fn{2~\MeV} \quad {\rm for} \ E_\nu<2\ \rm{MeV}\,,
  \label{eq:deltaphi}
  \end{equation} 
 Eq.~\eqref{eq:rate-separated-at 2MeV} becomes 
\al{
   {\df R\ov \df E^\pr}\fn{E^\pr} = 
   \int_0^{2\,\MeV}\df E_{\nu}\,\delta\Phi\fn{E_{\nu}}\,{\p\mc R\ov \p E^\pr}\fn{E_\nu, E^\pr}+h(E^\pr)\,,
\label{eq:integral-transform-deltaPhi}
}
which replaces Eq.~\eqref{eq:integral-transform} in our analysis.

Using Eqs.~\eqref{eq:rate-separated-at 2MeV}, \eqref{eq:h-definition} and \eqref{eq:integral-transform-deltaPhi}, the integrated rate over the $[E'_i,E'_{i+1}]$ bin is
\begin{eqnarray}
    R_i =\int_0^{2\,{\rm MeV}}dE_{\nu}~\delta\Phi(E_{\nu}) \mathcal{R}_i(E_{\nu})+h_i\,,
\end{eqnarray}
where we separated the contribution of high energy neutrinos (green dotted curve in 
Fig.~\ref{EventRate}):
\al{
    h_i &= \int_{E^\pr_i}^{E^\pr_{i+1}} \df E^\pr\, h\fn{E^\pr}\,.
}

\section{Numerical procedure}
\label{sec:numerical-procedure} 

\begin{figure} 
    \centering
\begin{tikzpicture}[scale = 0.98]
  \draw[-stealth] (-0.5,0) -- (7.5,0) node[right] {$E_\nu$};
  \draw[-stealth] (-0.5,0) -- (-0.5,4.5) node[above] {$\delta\Phi(E_\nu)$};

  \def\width{0.7}

  \def\hA{3.5}
  \def\hB{3.4}
  \def\hC{2.8}
  \def\hD{2.0}
  \def\hJ{1.5}
  \def\hN{1.0}

  \draw[thick,black] (0,0) -- (0,\hA);
  \draw[thick,black] (0,\hA) -- (\width,\hA) node[midway, above] {$\delta\Phi_1$};
  \draw[thick,black] (\width,\hA) -- (\width,0) node[below] {$E_\nu^2$};
  \draw[black] (0,0) -- (\width,0) node[midway, below left, black] {$E_{\nu}^\tx{min} = E_\nu^1$};

  \draw[thick,black] (\width,\hB) -- (2*\width,\hB) node[midway, above] {$\delta\Phi_2$};
  \draw[thick,black] (2*\width,\hB) -- (2*\width,0) node[below] {$E_\nu^3$};

  \draw[thick,black] (2*\width,\hC) -- (3*\width,\hC) node[midway, above] {$\delta\Phi_3$};
  \draw[thick,black] (3*\width,\hC) -- (3*\width,0) node[below] {$E_\nu^4$};

  \draw[thick,black] (3*\width,\hD) -- (4*\width,\hD) node[midway, above] {$\delta\Phi_4$};
  \draw[thick,black] (4*\width,\hD) -- (4*\width,0) node[below] {$E_\nu^5$};

  \fill (4.5*\width, 0.8) circle (1pt);
  \fill (4.7*\width, 0.8) circle (1pt);
  \fill (4.9*\width, 0.8) circle (1pt);

  \fill (6.8*\width, 0.8) circle (1pt);
  \fill (7.0*\width, 0.8) circle (1pt);
  \fill (7.2*\width, 0.8) circle (1pt);

  \draw[thick,black] (7.6*\width,0) node[below] {$E_\nu^{N_\tx{int}}$} -- (7.6*\width,\hN);
  \draw[thick,black] (7.6*\width,\hN) -- (8.6*\width,\hN) node[midway, above] {$\delta\Phi_{N_{\rm int}}$};
  \draw[thick,black] (8.6*\width,\hN) -- (8.6*\width,0) node[below right] {$E_\nu^{N_\tx{int} + 1}$};

  \draw[thick,black] (5.3*\width,0) node[below] {$E_\nu^j$} -- (5.3*\width,\hJ);
  \draw[thick,black] (5.3*\width,\hJ) -- (6.3*\width,\hJ) node[midway, above] {$\delta\Phi_j$};
  \draw[thick,black] (6.3*\width,\hJ) -- (6.3*\width,0);

\end{tikzpicture}
    \caption{Ansatz for the integrated antineutrino flux $\delta\Phi(E_\nu)$, Eq.~\eqref{StepForm}. The energy range of interest is divided into intervals of equal width, $j = 1, \ldots, N_{\rm int}$,  in which $\delta\Phi(E_\nu)$ takes a single value $\delta\Phi_j$. The $\delta\Phi_j$ are found by constrained minimization of the $\chi^2$ function in Eq.~\eqref{chiSq}. 
    }
    \label{neutrino-flux-hist}
\end{figure}

Following Ref.~\cite{Feldstein:2014gza}, we divide the neutrino energy range of interest $\sqbr{E_\nu^{\rm min}, 2~\MeV}$ into a large number $N_{\rm int}$ of intervals within which $\delta\Phi(E_\nu)$ takes a constant value $\delta\Phi_j$, 
\al{
\delta\Phi\fn{E_\nu} = \sum_{j=1}^{N_{\rm int}} \delta\Phi_j\,\sqbr{\Theta\fn{E_\nu - E_\nu^j} - \Theta\fn{E_\nu - E_\nu^{j+1}}}\,.
\label{StepForm}
}
For simplicity, here the intervals are chosen to have equal width, as shown in Fig.~\ref{neutrino-flux-hist} (though variable widths can also be implemented, as in Appendix~\ref{AppD:Method-1}).
In the best fit function $\delta\Phi^{\rm Best\,fit} (E_\nu)$,  many of the  $\delta\Phi^{\rm Best\,fit}_j$ in adjacent intervals turn out to be degenerate, producing a piecewise-constant function with at most $d-1$ steps, as in Eq.~\eqref{eq:FI-sum-deltas}. 

As described in Appendix~\ref{AppB:theoretical-nu-flux}, the theoretical modeling of the neutrinos flux changes at 2~MeV. The lower end  is fixed by the kinematic relation in Eq.~\eqref{maxRecoil}. Neglecting the energy resolution at threshold, i.e., taking $E_{\rm R}^{\rm min}=E^\pr_{\rm thr}$ as in Eqs.~\eqref{recoilRate1} 
and~\eqref{partialdcurlyH}, the minimum neutrino energy that produces a recoil above  threshold is $E_\nu^{\rm min} \simeq 0.18~\MeV$ for $E^\pr_{\rm thr} = 1~\eV$ and $E_\nu^{\rm min} \simeq 0.41~\MeV$ for $E^\pr_{\rm thr} = 5~\eV$.
 
\begin{figure}
     \centering
     \includegraphics[width=0.7\linewidth]{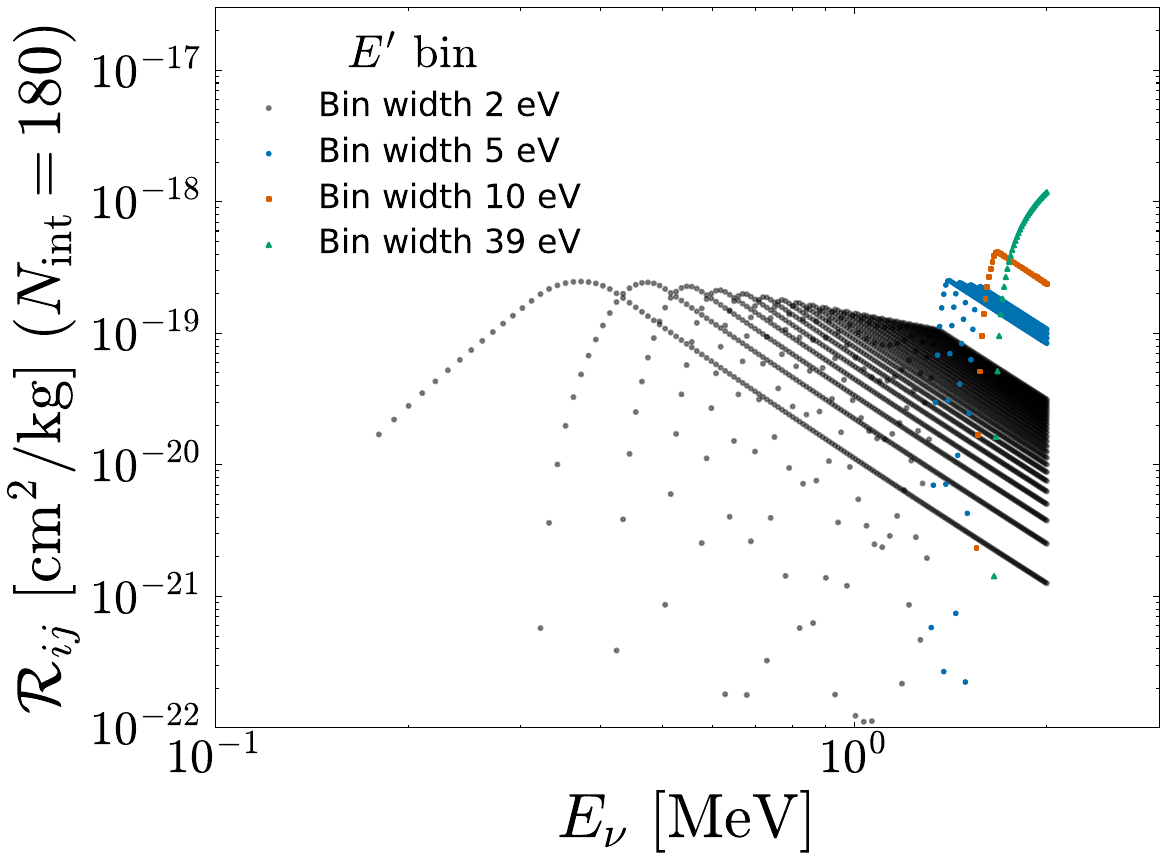}
     \caption{Response matrix elements $\mc R_{ij}$ for $N_{\rm int}=180$ intervals for $E_\nu <$ 2 MeV ($i= 1, \ldots , 31$ numbers the data bins, indicated by different symbols and colors, and $j= 1, \ldots, 180$ numbers the intervals in our ansatz). }
     \label{CRmat180_scatter}
 \end{figure}

With this ansatz for the integrated neutrino flux,  the rate in each of the $d$ bins, 
Eq.~\eqref{masterEquation},  becomes a linear combination of the parameters $\delta\Phi_j$, 
\al{
   R_i = R_{\sqbr{E_i^\pr, E_{i + 1}^\pr}} = \sum_{j = 1}^{N_\tx{int}} \mc R_{ij}\,\delta\Phi_j + h_i, 
    \label{eq:predicted-rate}
    }
with coefficients $R_{ij}$ determined by integrating the response function of each data bin $i$, $R_{\sqbr{E_i^\pr, E_{i + 1}^\pr}}\fn{E_\nu}$, 
in each $E_\nu$ interval $j$:
    \al{
    \mc R_{ij} \equiv \int_{E_\nu^j}^{E_\nu^{j+1}} \df E_\nu\, \mc R_{\sqbr{E_i^\pr, E_{i + 1}^\pr}}\fn{E_\nu}.
     \label{eq:predicted-rate-coeff}
}
 We evaluate these response matrix elements $\mc R_{ij}$ numerically. An example for $N_{\rm int}= 180$ below $E_\nu = 2$~MeV is shown in Fig.~\ref{CRmat180_scatter}.

 With our ansatz for the integrated neutrino flux, the  predicted number of events in the $i^{\rm th}$ data bin is
\al{
    N_i = \mc E\,\pn{\sum_{j = 1}^{N_\tx{int}} \mc R_{ij}\,\delta\Phi_j + h_i + B_i}\,,
    \label{eq:Ni-predicted}
}
where $B_i$ are the number of background events per kg$\cdot$year. 
We consider three possibilities for $B_i$.
One is the unrealistic assumption of no background (labeled ``No Bkg'' in the figures) with $B_i=0$. 
Another is a relatively large exponential background (labeled ``Exp Bkg''), which, following Ref.~\cite{NUCLEUS:2019igx} we parameterize as
\al{
    B\fn{E^\pr} = A\,e^{-E^\pr/D} + C\,.
\label{bkg-form}
}
Then in each bin,
\al{
    B_i \equiv B_{\sqbr{E_i^\pr, E_{i + 1}^\pr}} = \int_{E^\pr_i}^{E^\pr_{i + 1}} \df E^\pr\, B\fn{E^\pr}\,.
    \label{bkg}
}
We choose $A=$ 460$/$(keV$\cdot$kg$\cdot$day), $D=$ 10~eV and $C=$ 100$/$(keV$\cdot$kg$\cdot$day), so that the background and $\CEnNS$ signal events in the lowest energy bin are equal; see Fig.~\ref{EventRate}.
Finally, to compare with ``scenario 2''  of Ref.~\cite{Liao:2023kyy}, we adopt a flat background with $B\fn{E^\pr} =$ 1$/$(keV$\cdot$kg$\cdot$day) (labeled ``Flat Bkg''). 

 The $\{\delta\Phi_j\}$ parameters are determined by minimizing the Neyman $\chi^2$ statistic,
\al{
    \chi^2 = \sum_i {1\over \mc E\mc O_i} \sqbr{\mc E\pn{\displaystyle\sum_{j = 1}^{N_\tx{int}}\mathcal R_{ij}\delta\Phi_j + h_i + B_i} - \mc E\mc O_i}^2\,,
\label{chiSq}
}
subject to the nonnegativity and nonincreasing (monotonicity) constraints,
\al{
    \delta\Phi_j \geq 0\,, \qquad \delta\Phi_{j+1} \leq \delta\Phi_j\,, \qquad j = 1, \ldots, N_{\rm int} - 1\,.
\label{eq:constraints}
}

In Eq.~\eqref{chiSq} $\mc O_i$ is the per-unit-mass mock ``observed'' event rate in the $i^{\rm th}$ bin,
and $\mc E \mc O_i$ is the corresponding number of events observed with exposure $\mc E$. 
 
\begin{figure} 
    \centering
    \includegraphics[width=0.7\linewidth]{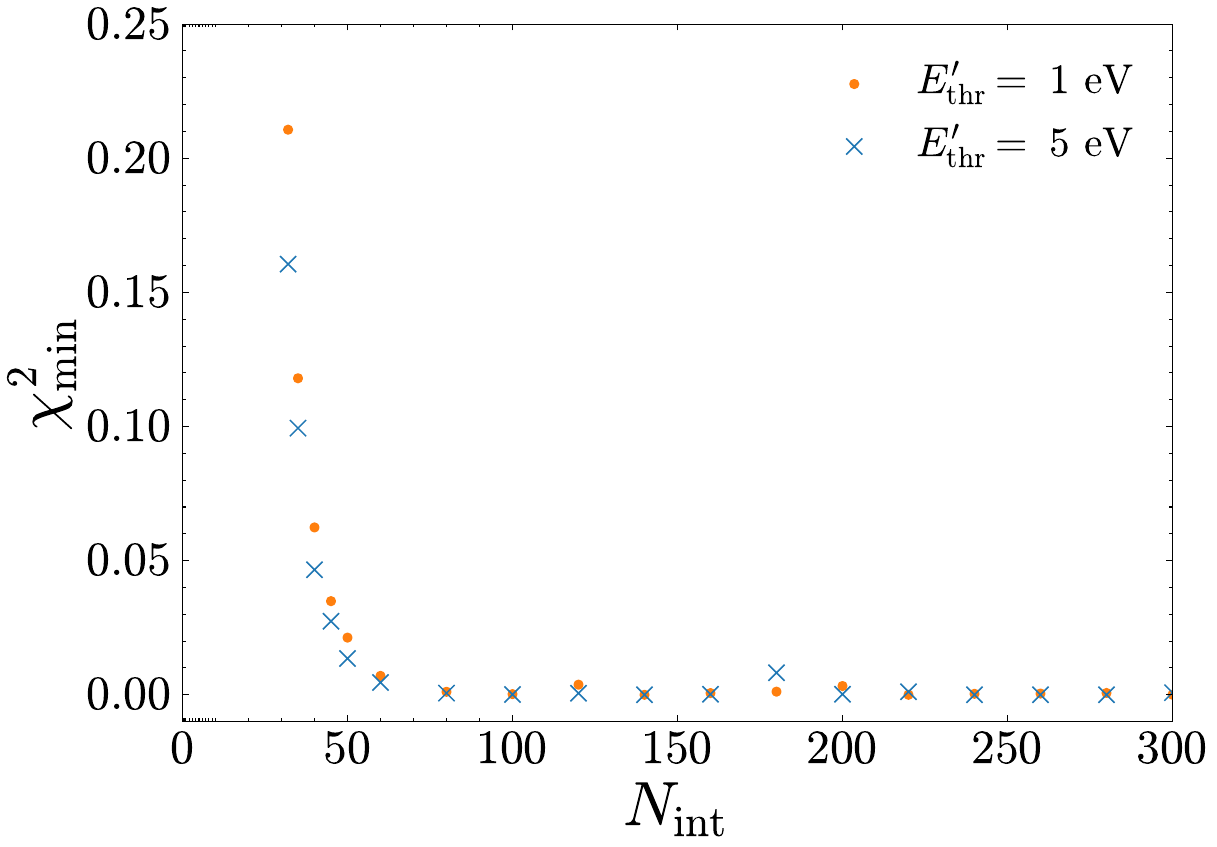}
    \caption{
    Minimum Neyman $\chi^2$ values, $\chi^2_{\min}$, as a function of the number of intervals  $N_{\rm int}$ in our ansatz, assuming an exposure $\mc E= 3$~kg$\cdot$year, no background,  thresholds 
    $E^\pr_{\rm thr} = 1$~eV (orange dots) and 5~eV (blue crosses), and $d=31$ and 29 data bins, respectively. 
    Clearly, $\chi^2_{\min}$ decreases very fast with $N_{\rm int}$ until it plateaus for $N_{\rm int}\gg d$. 
    }
    \label{chiSquaredComparison}
\end{figure}

Recall that we choose $N_{\rm int}$ by requiring that increasing it further no longer improves the fit. Two examples of how the minimum value $\chi^2_{\min}$ changes with $N_{\rm int}$ is shown in Fig.~\ref{chiSquaredComparison} assuming $B_i=0$, $\mc E =3~\kg\cdot$year and $E^\pr_{\rm thr} = 1~\eV$ (orange dots) or $5~\eV$ (blue crosses), with $d=31$ and 29, respectively.  It is evident that  $\chi^2_{\min}$ decreases rapidly with $N_{\rm int}$ until it plateaus at a value much larger than the number of data points. For $N_{\rm int} \lesssim 50$ the fit is still poor. As  $N_{\rm int}$ approaches 180, the piecewise constant shape of the best-fit function becomes stable -- a further increase in $N_{\rm int}$ does not change it -- indicating that the downward steps have been adequately located.
 We use $N_{\rm int} = 180$ to find the best-fit functions and $N_{\rm int} = 80$ to find the confidence bands (see Appendix~\ref{AppC:Monte-Carlo}).

As explained in Section~\ref{Sec:our-method}, to obtain the pointwise confidence band at each energy $E_\nu$, a new best-fit function is found while fixing the integrated rate at the particular energy $\delta\Phi(E_\nu)$ above or below the ``global'' best-fit. The new fit necessarily has a larger minimum $\chi^2$ value than the original fit and the profile likelihood ratio $\Delta L_{\rm min}$ is the difference between the two, 
$\Delta L_{\rm min}= \Delta \chi^2_{\rm min}$.  
To compute the band at the energy $E_\nu^j$ corresponding to the $j^{\rm th}$ interval, we fix the parameter $\delta\Phi_j$ to say 
$\delta\Phi_j^\ast$, different from its global best-fit value, and minimize Eq.~\eqref{chiSq} over the remaining  $\{\delta\Phi_k\}_{k \neq j}$, which are effectively nuisance parameters, retaining throughout the constraints in Eq.~\eqref{eq:constraints}.
The result is the profile $\chi^2$ as a function of the fixed value,
\al{
    \Delta\chi^2_{\min}(\delta\Phi_j^\ast) \equiv \chi^2_{\min}(\delta\Phi_j = \delta\Phi_j^\ast) - \chi^2_{\min}\,,
\label{eq:Delta-chisq}
}
where $\chi^2_{\min}$ is the unconstrained global minimum.
For trial values $\delta\Phi_j^\ast$ close to the best fit, we find that $\Delta\chi^2_{\min}(\delta\Phi_j^\ast)$ is numerically indistinguishable from zero. This signals the existence of a degeneracy band~\cite{Gelmini:2017aqe}. The best-fit function is nonunique. In the following figures, we indicate the region in which $\Delta\chi^2_{\min} \leq 10^{-3}$, an upper limit somewhat arbitrarily chosen, and well below the value of any meaningful confidence level.

 Were the best-fit function unique, $\Delta\chi^2_{\min}(\delta\Phi_j^\ast)$ by Wilks' theorem would follow a $\chi^2$ distribution with one degree of freedom, and the attribution of a confidence level would be analytic~\cite{Gelmini:2017aqe}. The presence of the degeneracy band invalidates Wilks' theorem. We therefore
 find  the correspondence between $\Delta\chi_{\min}^2$ and confidence levels using Monte Carlo techniques via a parametric bootstrap (see e.g. Ref.~\cite{Demortier:2007zz} and 
 Section~13 of Ref.~\cite{Cousins:2018tiz}). We describe our procedure in Appendix~\ref{AppC:Monte-Carlo}.
  
 \section{Results}
\label{Sec:Results}
\begin{figure}
\centering
\begin{tabular}{cc}
    \begin{minipage}{0.5\columnwidth}
    \includegraphics[height = 0.70\textwidth]{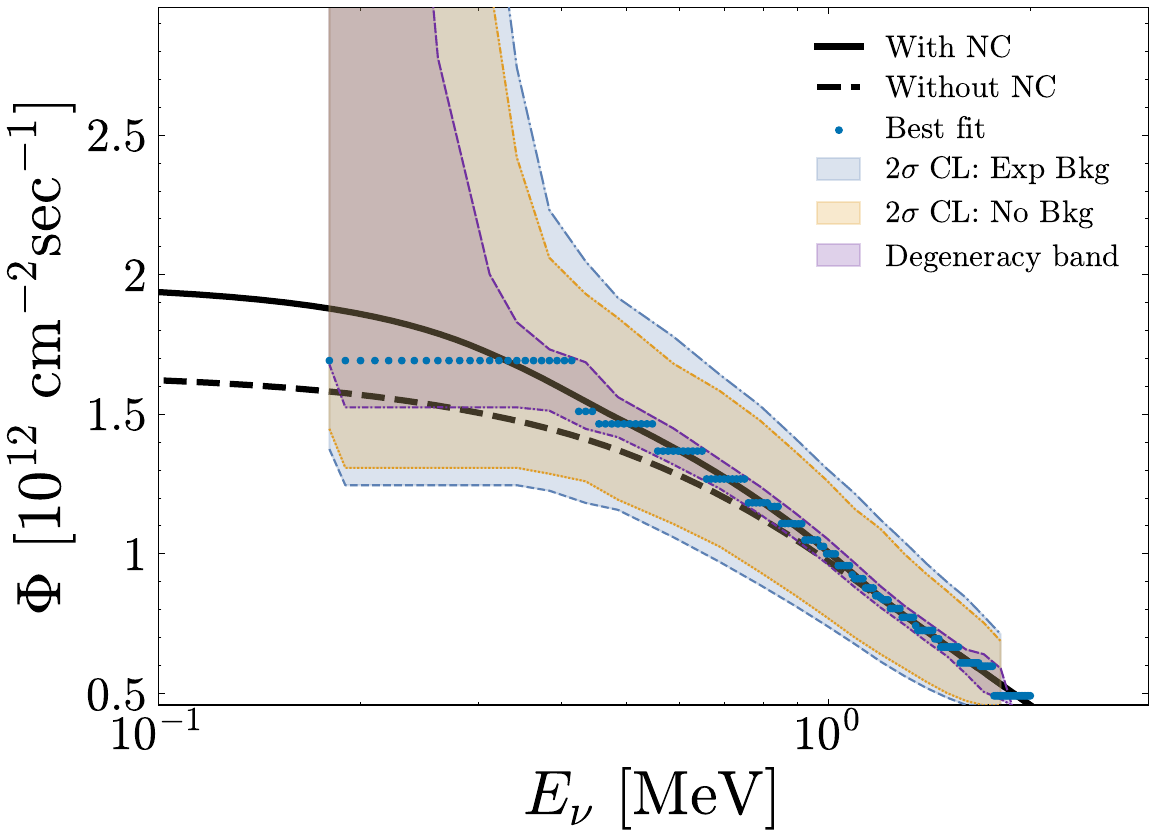}
  \end{minipage}
    \begin{minipage}{0.5\columnwidth}
    \includegraphics[height = 0.70\textwidth]{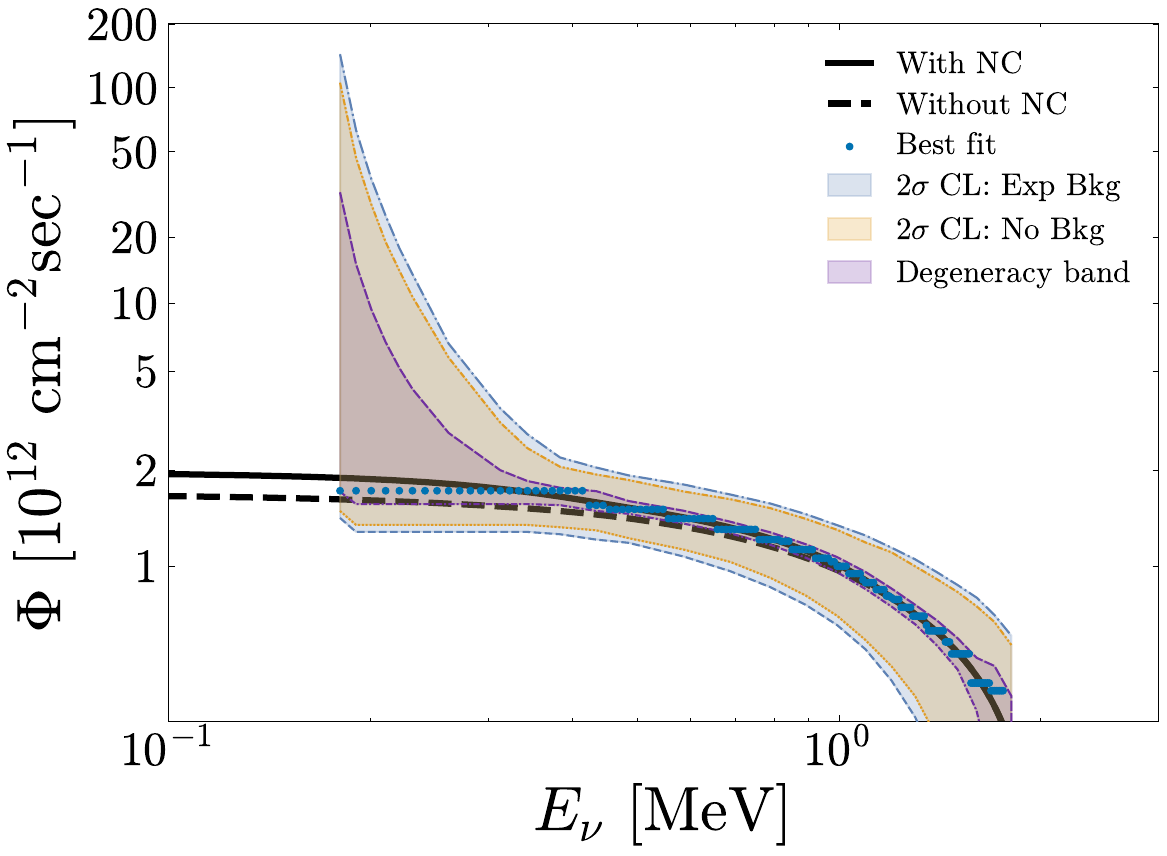}
  \end{minipage}
\end{tabular}
\begin{tabular}{cc}
    \begin{minipage}{0.5\columnwidth}
    \includegraphics[height = 0.70\textwidth]{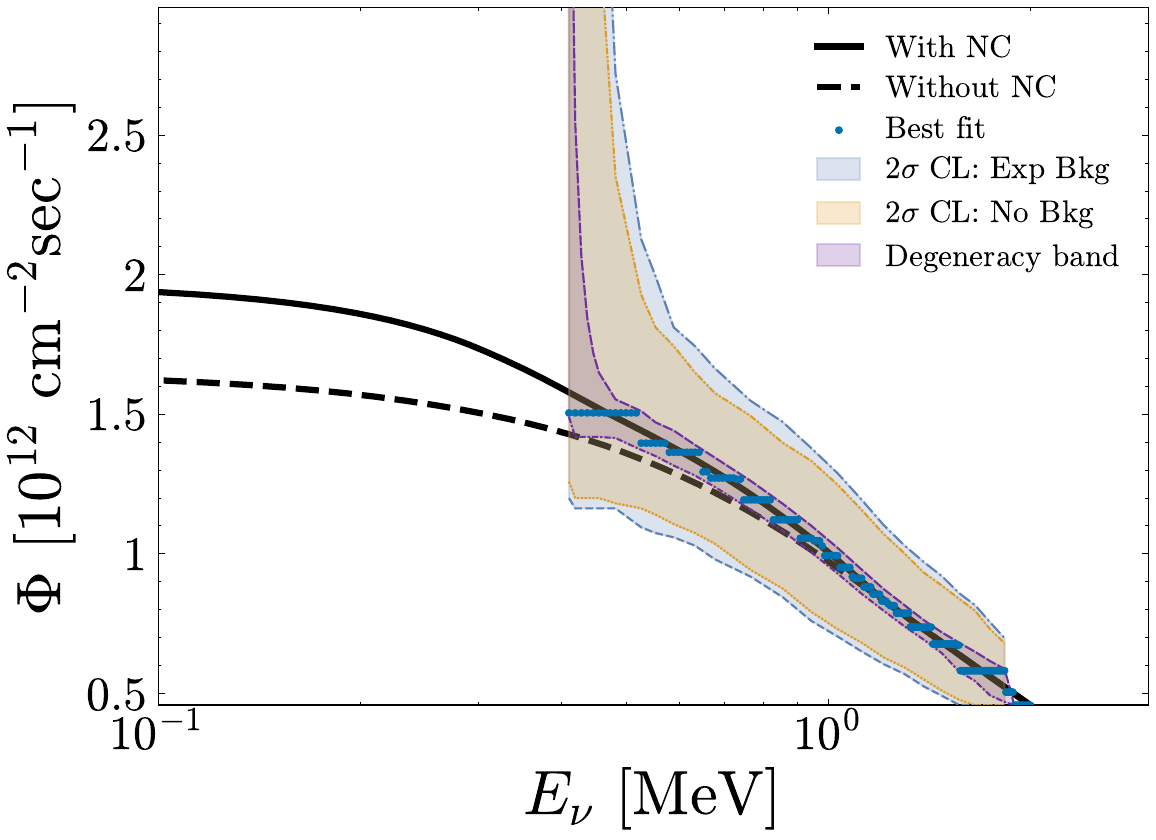}
  \end{minipage}
    \begin{minipage}{0.5\columnwidth}
    \includegraphics[height = 0.70\textwidth]{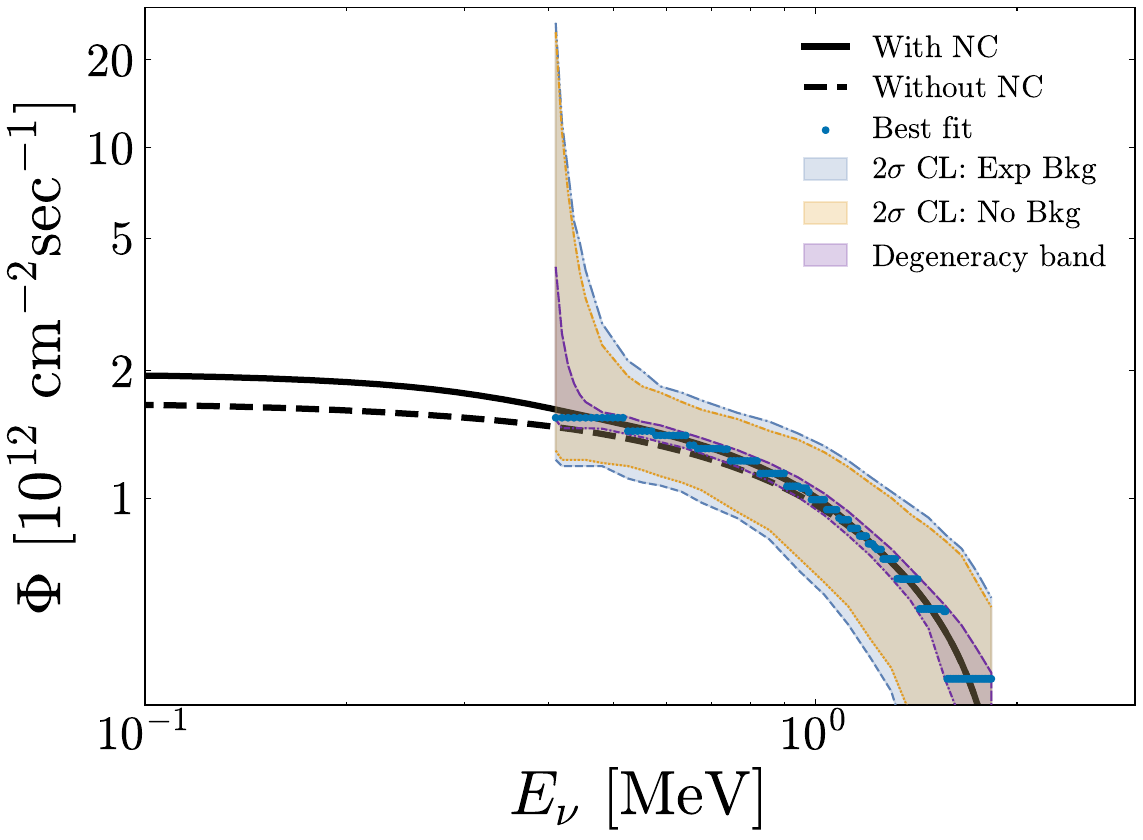}
  \end{minipage}
\end{tabular}
    \caption{Piecewise constant best-fit integrated reactor antineutrino flux $\Phi = \delta\Phi + \Phi\fn{2~\MeV}$ (blue dots), pointwise degeneracy band (purple) and 2$\sigma$ pointwise confidence bands, up to 2~MeV, assuming an exposure of $3~\kg\cdot$year and either no background (light brown) or the exponential background of Eqs.~\eqref{bkg-form} and~\eqref{bkg} (light blue) and $E^\pr_{\rm thr} = 1~\eV$ (top panels) or $5~\eV$ (bottom panels), on a linear scale (left panels) or a logarithmic scale (right panels). The best-fit is the same for both backgrounds. Only the degeneracy band for the exponential background is shown because the band for the no-background case is almost identical. 
    The black solid and dashed curves are the theoretical fluxes in Fig.~\ref{fig-Int-Flux}. 
   }
    \label{OnlyHighEnergyNeutrino}
\end{figure}

Our main results for the unfolded integrated neutrino flux $\Phi (E_\nu) = \delta\Phi (E_\nu) + \Phi\fn{2~\MeV}$ for $E_\nu<2$ MeV are shown in Fig.~\ref{OnlyHighEnergyNeutrino}. The flux 
$\Phi$ is on a linear scale in the left panels and on a logarithmic scale in the right panels.  The  2$\sigma$ pointwise confidence bands are shown assuming an exposure of $3~\kg\cdot$year and no background $B_i=0$ (light brown), or the exponential background of Eqs.~\eqref{bkg-form} and~\eqref{bkg} (light blue), for threshold energies of  $E^\pr_{\rm thr} = 1~\eV$ (top panels) or $5~\eV$ (lower panels).  The best fit (shown with blue dots, each corresponding to one of the $N_{\rm int}$ intervals in Eq.~\eqref{StepForm}) is the same independently of the background. It is a piecewise constant function with less than $d-1$ downward steps, 27 for the 1~eV threshold (which has $d=31$ data points) and 24 for the 5~eV threshold (with $d=29$). The  pointwise degeneracy band (purple) with $\Delta\chi^2_{\min} < 10^{-3}$, is shown only for the exponential background, since the band for the case with no background is practically identical. 
The black solid and dashed curves are the theoretical fluxes in Fig.~\ref{fig-Int-Flux}.
 Both fluxes are well within the confidence bands and are not distinguishable. As we show below, a much larger exposure is needed to distinguish between them.

\begin{figure}
    \centering
    \includegraphics[width=0.485\linewidth]{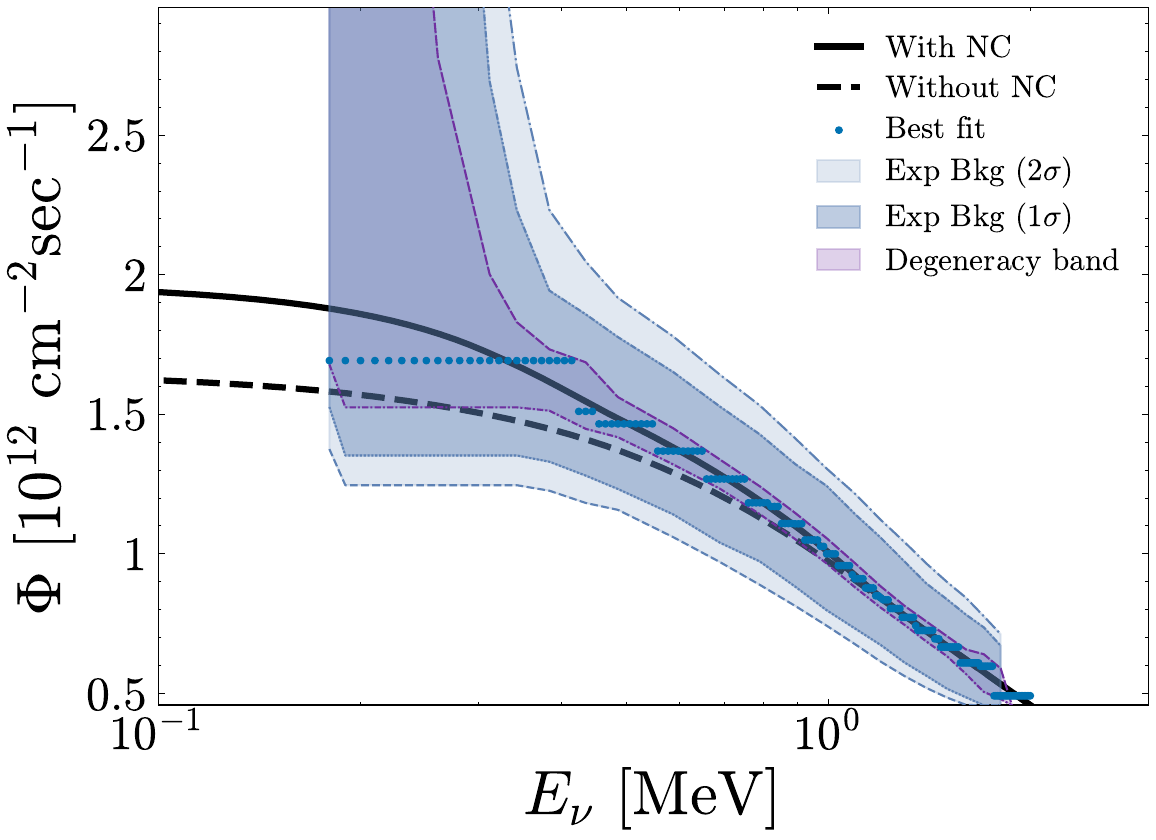}
    \includegraphics[width=0.49\linewidth]{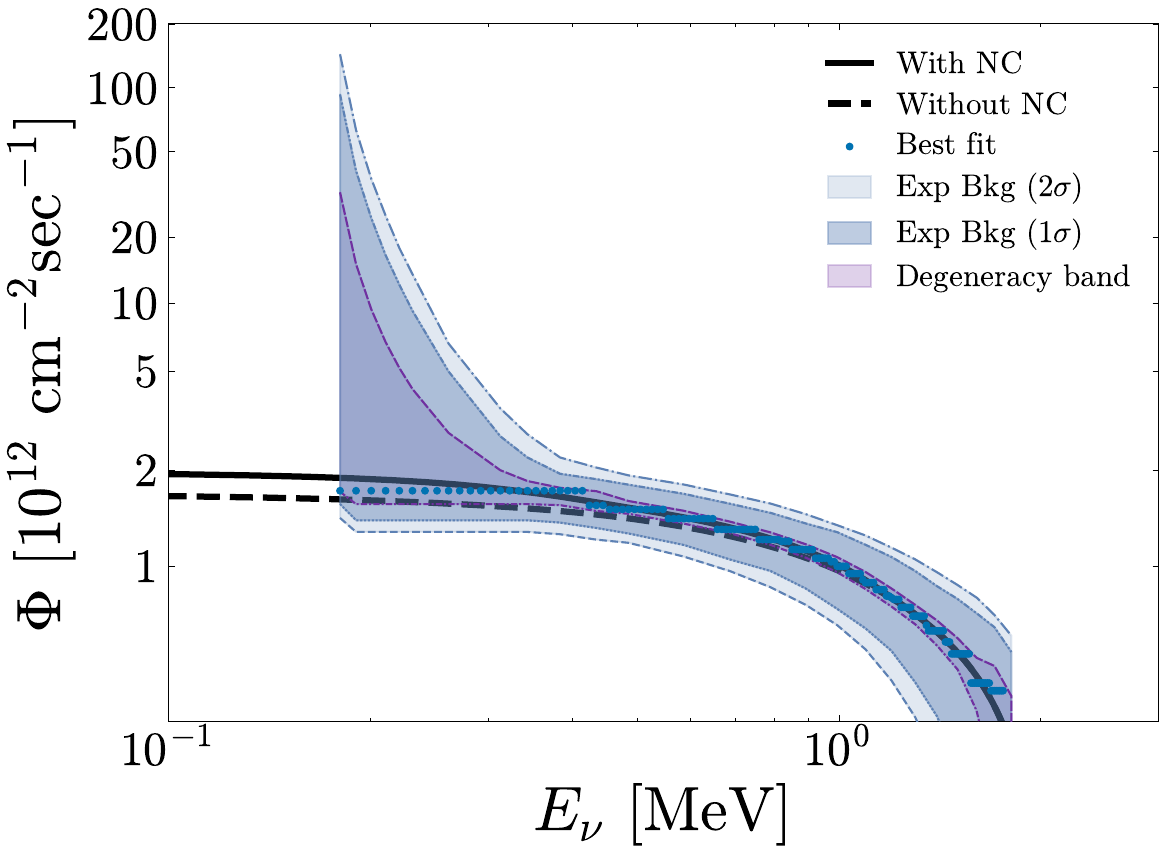}
    \caption{Same as Fig.~\ref{OnlyHighEnergyNeutrino}, but showing the $1\sigma$ (dark) and $2\sigma$ (light) CL bands assuming the exponential background, for $E^\pr_{\rm thr}=1$~eV and $\mc E = 3$~kg$\cdot$yr.}
    \label{1sigma2sigma}
\end{figure}

The bands at low energy are extremely wide primarily because their upper edge extends substantially further than their lower edge. This is evident in the 
right panels, where $\Phi$ is plotted on a logarithmic scale, and can be understood as follows. When we fix one of the $\delta\Phi_j$ in Eq.~\eqref{StepForm} to a value $\delta\Phi_j^\ast$ above the best fit (i.e., above the blue dot at the particular energy in Fig.~\ref{OnlyHighEnergyNeutrino}) and proceed to find the new best fit, due to the monotonicity constraint in Eq.~\eqref{eq:constraints}, all the $\delta\Phi_k$ values for $k<j$ are required to also become at least as large as their best-fit values. This increases the minimum $\chi^2$ with respect to its value corresponding to the best fit indicated by the blue dots.  On the other hand, for $k>j$, the values $\delta\Phi_k$ can continue to remain at their best-fit values, thus not increasing the minimum $\chi^2$ of the new fit. Therefore, when $j$ is close to unity, only a few of the lowest energy $\delta\Phi_k$ contribute to changes in the constrained minimum $\chi^2$ away from its global minimum. Consequently, the $\delta\Phi_k$ values can be very large without producing a significant change in $\Delta \chi^2_{\min}$. This does not happen when $\delta\Phi_j^\ast$ is below the best fit. Many of the $\delta\Phi_k$ values with $k>j$ must decrease to be at most as large as $\delta\Phi_j^\ast$, which rapidly increases the minimum $\chi^2$ value, causing the lower edge of the bands to not extend far from the blue dots.

We consider the no-background case as the extreme limit of a very small background, while the exponential background accounts for the opposite case of a large background equal to the total $\CEnNS$ rate in the lowest energy bin. The confidence bands for the two background models are very similar, showing that our reconstruction method is largely insensitive to backgrounds. Although we are convinced that this result is a characteristic of our method, understanding the underlying reason is left for future work.

Because the number of mock events per experimental bin is computed by summing the background $B_i$ in each bin, the numerators of all the terms in the $\chi^2$ function, Eq.~\eqref{chiSq}, do not change with the choice of background model.  This leads to an identical best-fit function independently of the background model (although the denominators of each term depend on the background). Because the signal and backgrounds are proportional to the exposure, the $\chi^2$ is also proportional to the exposure, so that the minimum $\chi^2$ of the fits changes with exposure but not the best-fit functions. Thus, our best-fit functions do not change with the background model or exposure, but only with the energy threshold.

In Fig.~\ref{1sigma2sigma} we show the 1$\sigma$ and 2$\sigma$ bands assuming the exponential background, for the experimentl threshold $E^\pr_{\rm thr}=1$~eV and exposure $\mc E = 3$~kg$\cdot$yr.

\begin{figure}
\centering
\begin{tabular}{cc}
    \begin{minipage}{0.5\columnwidth}
    \includegraphics[height = 0.7\textwidth]{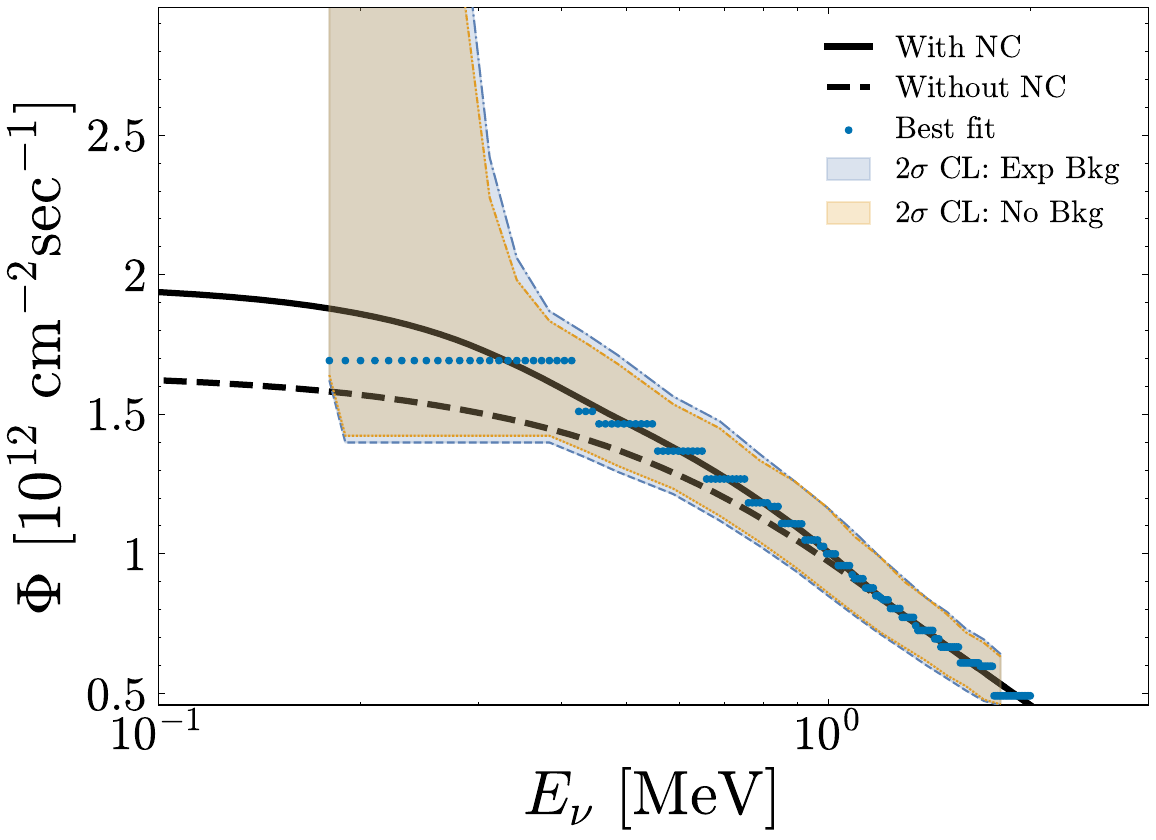}
  \end{minipage}
    \begin{minipage}{0.5\columnwidth}
    \includegraphics[height = 0.7\textwidth]{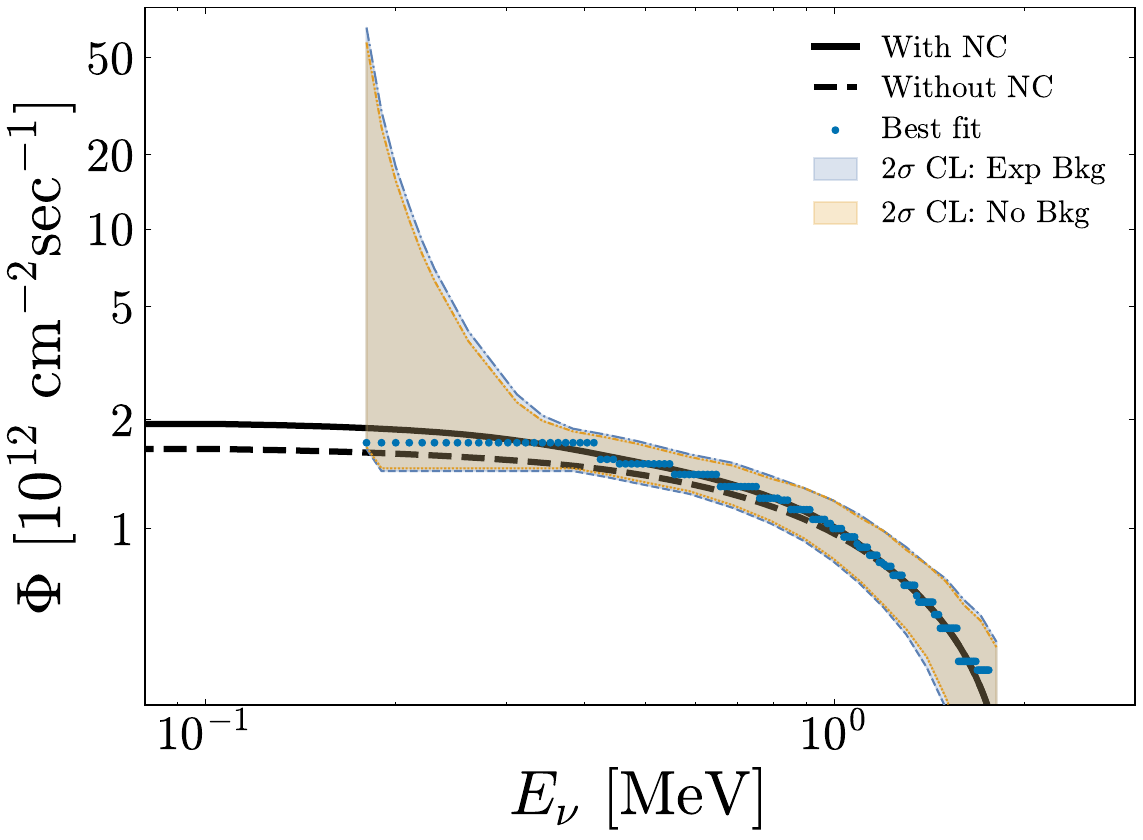}
  \end{minipage}
\end{tabular}
    \caption{Similar to Fig.~\ref{OnlyHighEnergyNeutrino} but for $E^\pr_{\rm thr} = 1~\eV$ and $\mc E=50$ kg$\cdot$year, the minimum exposure 
    for which the dashed curve is close to the lower-boundary of the 2$\sigma$ band at some energy.  
    }
    \label{1eV 50kgyrs}
\end{figure}

\begin{figure}
\centering
\begin{tabular}{cc}
    \begin{minipage}{0.5\columnwidth}
    \includegraphics[height = 0.7\textwidth]{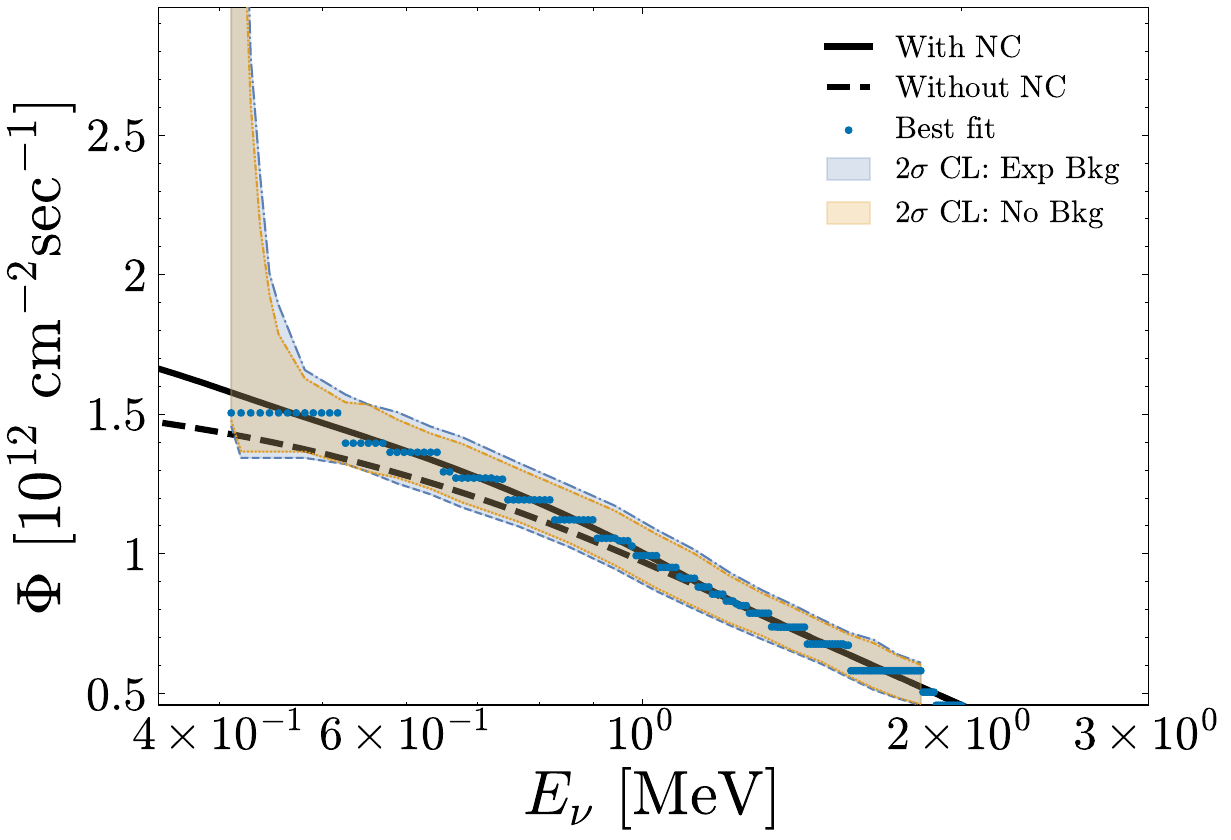}
  \end{minipage}
    \begin{minipage}{0.5\columnwidth}
    \includegraphics[height = 0.7\textwidth]{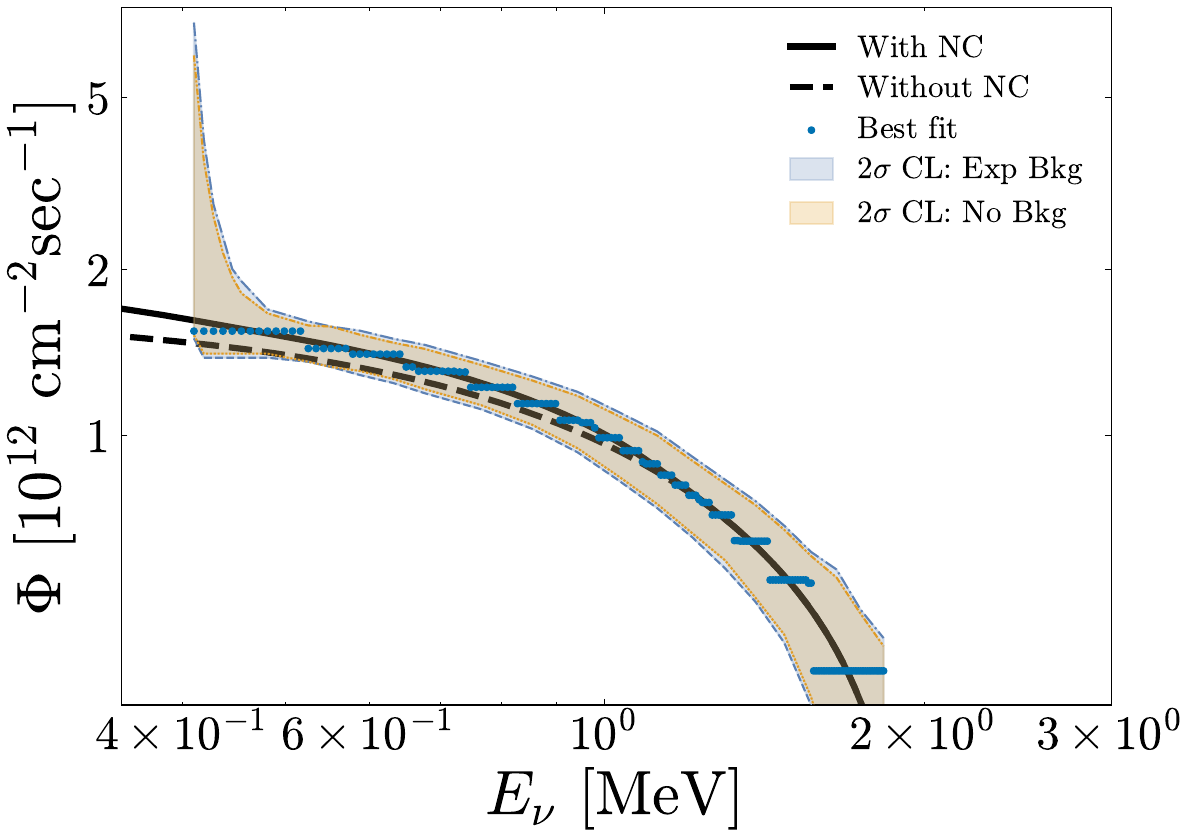}
  \end{minipage}
\end{tabular}
    \caption{Similar to Fig.~\ref{OnlyHighEnergyNeutrino} but for $E^\pr_{\rm thr} = 5~\eV$ and
    $\mc E=330$ kg$\cdot$year, the minimum exposure 
    for which the dashed curve is close to the lower-boundary of the 2$\sigma$ band at some energy.    }
    \label{5eV 330kgyrs}
\end{figure}

Our mock data were produced with the theoretical model  that includes the neutron-capture component, indicated by the black solid curve in Fig.~\ref{OnlyHighEnergyNeutrino}, and the piecewise constant best-fit functions trace it rather well. However, the confidence bands obtained with an exposure of 3~kg$\cdot$year do not allow us to distinguish this flux from one that excludes the neutron-capture component, shown by the black dashed curve. In Figs.~\ref{1eV 50kgyrs} and~\ref{5eV 330kgyrs}  exposures of 50~kg$\cdot$year and 330~kg$\cdot$year are assumed, with 1~eV and 5~eV thresholds, respectively. The dashed curve  almost touches the 2$\sigma$ band at some energy,  indicating that a larger exposure is necessary for parts of the curve to lie outside the band.
A lower threshold clearly improves discrimination, as shown by the roughly one order of magnitude smaller exposure required at 1~eV than at 5~eV. However, establishing the existence of the neutron-capture component requires exposures that are prohibitively large and unrealistic for the foreseeable future. This aligns with the conclusion reached in Ref.~\cite{Liao:2023kyy} using Tikhonov-regularized unfolding.

\begin{figure}
\centering
\begin{tabular}{cc}
    \begin{minipage}{0.5\columnwidth}
    \includegraphics[height = 0.7\textwidth]{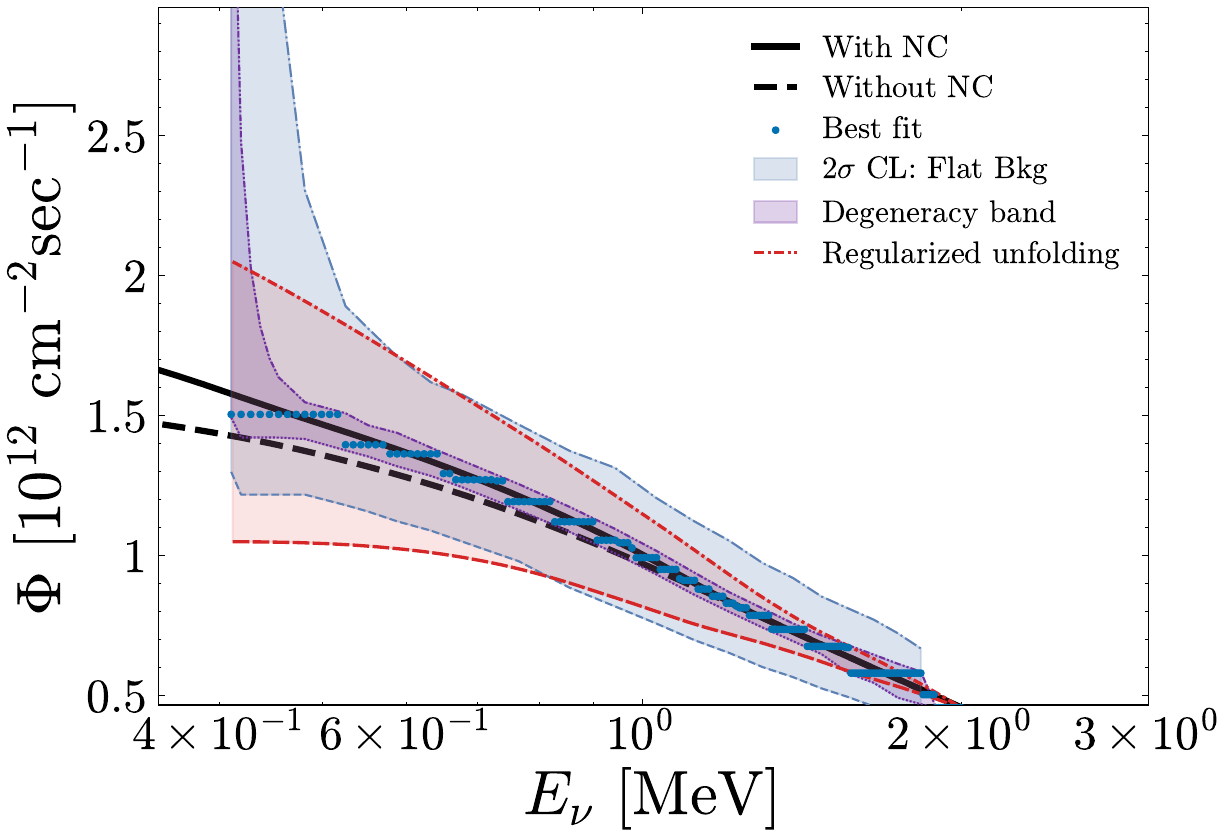}
  \end{minipage}
    \begin{minipage}{0.5\columnwidth}
    \includegraphics[height = 0.7\textwidth]{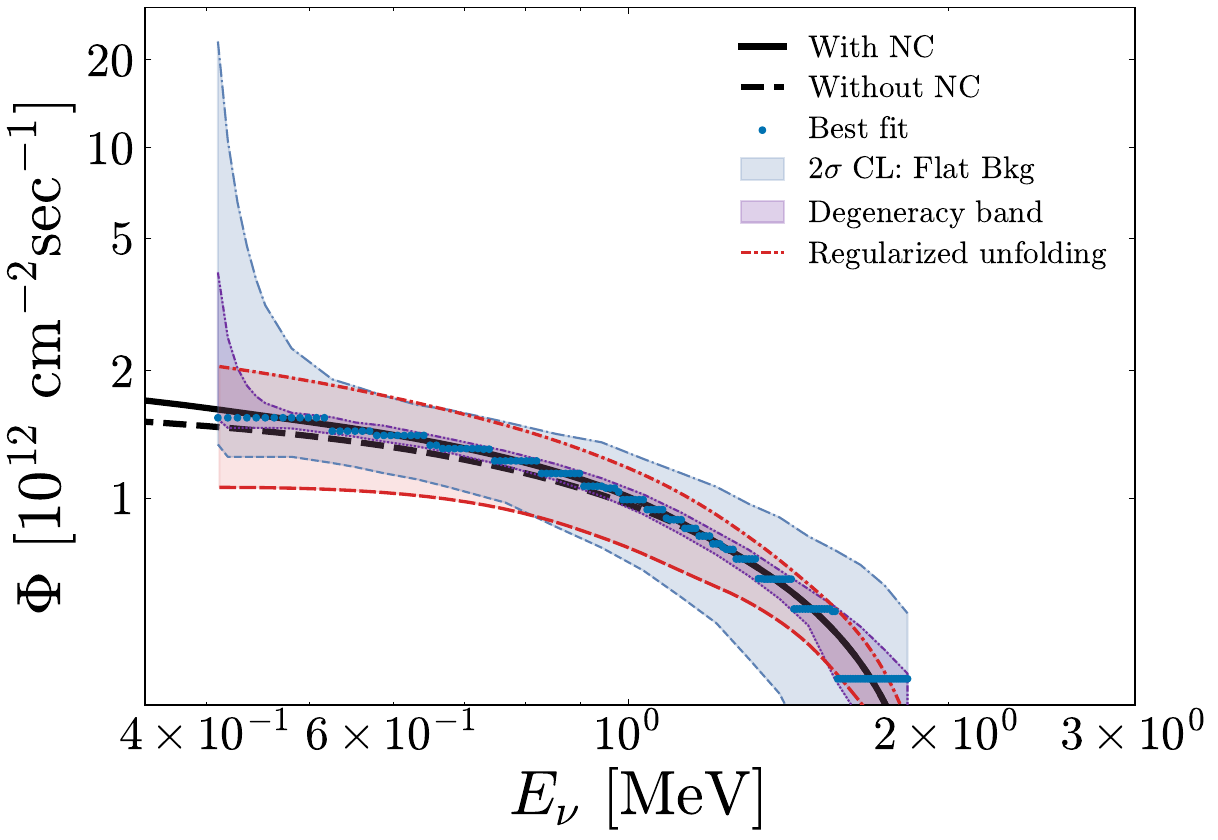}
  \end{minipage}
\end{tabular}
    \caption{Comparison of the pointwise 2$\sigma$ band for the integrated neutrino flux $\Phi = \delta\Phi + \Phi\fn{2~\MeV}$ obtained with our method (light blue) and the integrated boundaries of the 2$\sigma$ band in differential flux obtained by Tikhonov-regularized unfolding in Ref.~\cite{Liao:2023kyy} (pink), assuming the parameters of their ``scenario~2'', namely  a experimental threshold $E^\pr_{\rm thr} = 5~\eV$,  3~kg year of exposure, and a flat background of 1$/$(keV$\cdot$kg$\cdot$day). Linear scale (logarithmic) scale  is used in the left (right) panel on the vertical axis. The (background-independent) best-fit (blue dots) and degeneracy band (purple) for the flat background obtained with our method are also shown.
    }
    \label{Flat Bkg 5eV 3kgyrs}
\end{figure}

To compare our unfolding method with the method of Ref.~\cite{Liao:2023kyy}, we obtain with our method the 2$\sigma$ pointwise confidence band for ``scenario~2'' of Ref.~\cite{Liao:2023kyy} with $E^\pr_{\rm thr} = 5~\eV$,  $\mc E = 3~\kg\cdot$year and a flat background of 1$/$(keV$\cdot$kg$\cdot$day).
In constructing our band we did not fluctuate the $h_i$ and $B_i$ contributions independently, to match the procedure of Ref.~\cite{Liao:2023kyy}.
The band of Ref.~\cite{Liao:2023kyy} is obtained by integrating the boundaries of the 2$\sigma$ pointwise confidence band on the differential neutrino flux, in the left panel of their Fig.~7.
The two bands are shown in Fig.~\ref{Flat Bkg 5eV 3kgyrs}.
 Our band (light blue) and the integrated band of Ref.~\cite{Liao:2023kyy} (pink) are shown in linear (left panel) and logarithmic (right panel) scales on the vertical axis. Except for the upper edge at low energies where our band extends to very high $\Phi$ values and is nonconstraining, the two bands are comparable. The  lower edge of our band is tighter than theirs for $E_\nu \lesssim 0.8~\MeV$ and vice-versa for $E_\nu \gtrsim 0.8~\MeV$.

The integration of the boundaries of a band preserves the confidence level only if the errors are completely positively correlated. This is easy to prove by propagating the errors in two bins to their sum, in which case the errors just sum if they have a perfect positive correlation i.e., $(A \pm \sigma_A) + (B \pm \sigma_B)= A +B \pm (\sigma_A + \sigma_B)$. The error in the sum is maximum. A regularized framework, such as that of Ref.~\cite{Liao:2023kyy}, produces very positively correlated errors -- the stronger the regularization the higher the correlation of errors -- although inasmuch as the correlation is not perfect, the confidence level in the integrated flux band can only increase. Since the actual combined error is smaller without perfect positive error correlation, the integrated band is wider and more conservative than a true 2$\sigma$ confidence interval. 

\begin{figure*}
    \centering
    \includegraphics
    [width=0.9\linewidth]{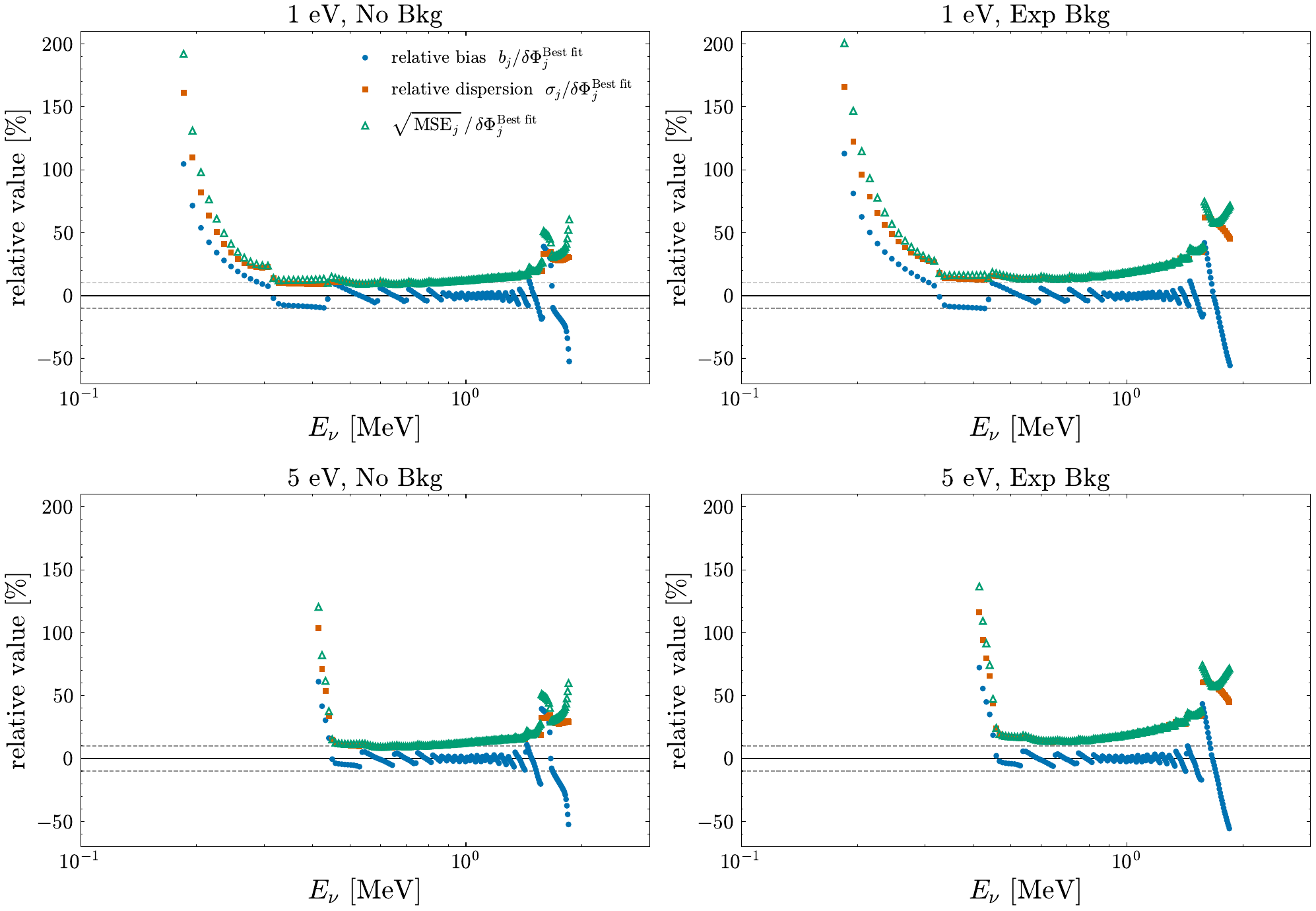}
    \caption{Bias (blue dots), dispersion (red squares) and $\sqrt{\rm MSE}$ (green triangles) relative to the best-fit function as a function of $E_\nu$, for a 1~eV (upper panels) and 5~eV (lower panels) threshold, a 3 kg$\cdot$year  exposure, and $N_{\rm int} = 180$. Left panels assume no background, while right panels include the exponential background. The horizontal lines mark $0$ and $\pm 10$\% to guide the eye.
     }
    \label{fig:Relative-b-sigma-sqrtMSE}
\end{figure*}

\begin{figure*}
    \centering
    \includegraphics
    [width=0.9\linewidth]{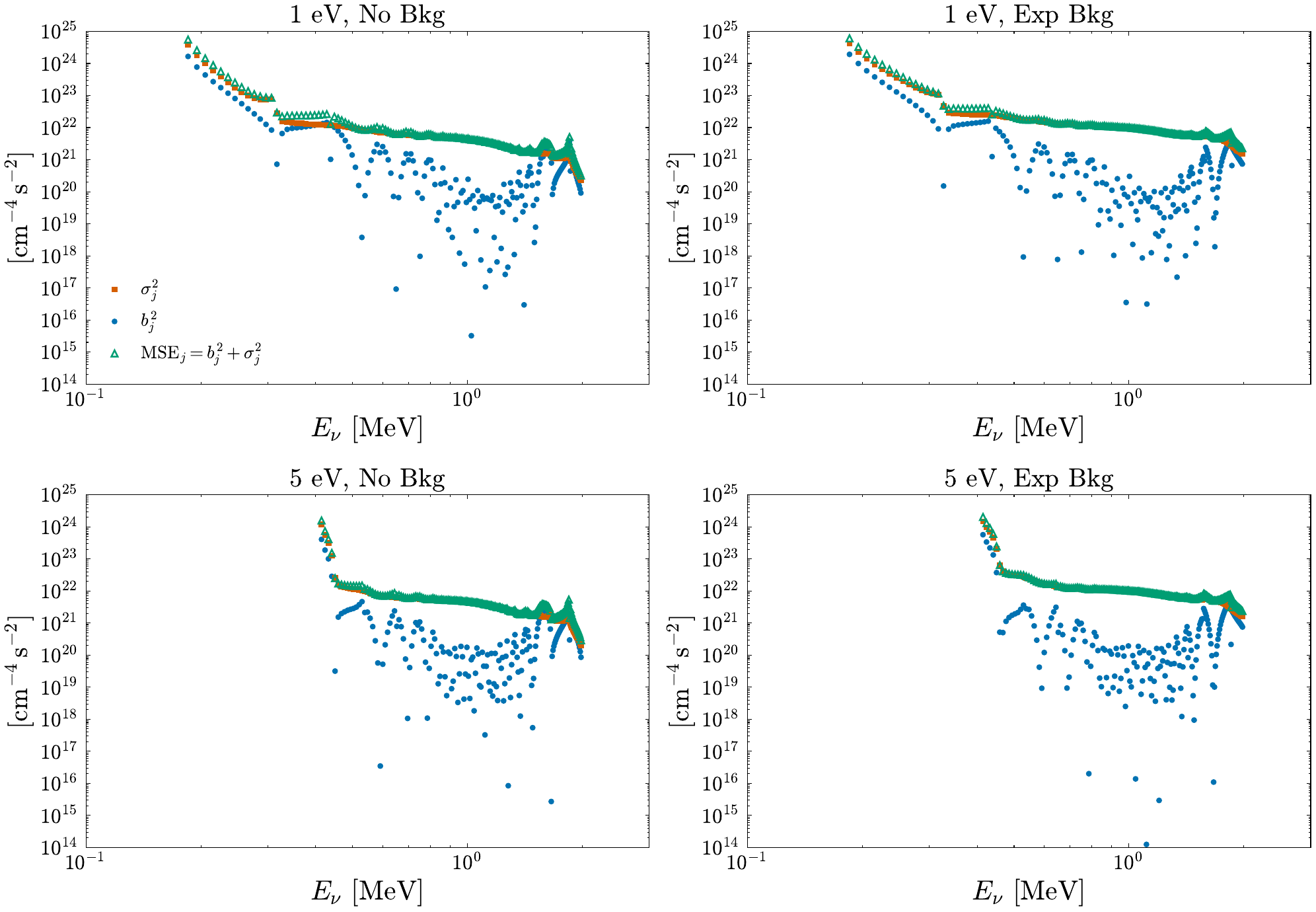}
    \caption{Bias-squared (blue dots), variance (red squares) and  MSE (green triangles) as a function of $E_\nu$, for a threshold of 1~eV(upper panels) and 5~eV (lower panels), a  3 kg~yrs exposure, $N_{\rm int} = 180$ and  either no background (left panels) or exponential background (right panels).
  }
    \label{fig:b2-variance-MSE}
\end{figure*}

\begin{figure*}
    \centering
    \includegraphics
    [width=0.9\linewidth]{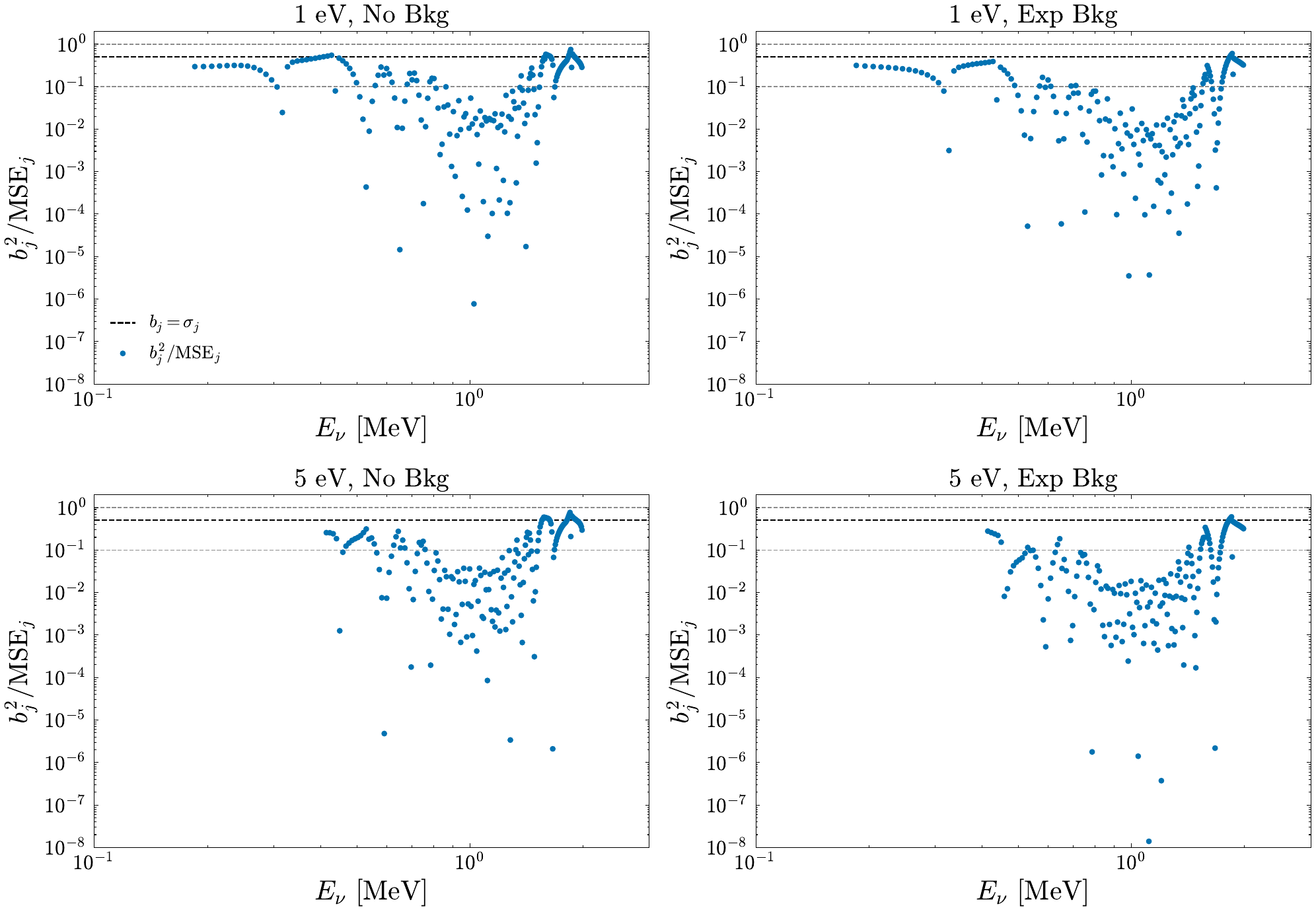}
    \caption{Ratio of bias-squared to MSE as a function of $E_\nu$, 
    for a 1~eV (upper panels) and 5~eV (lower panels) threshold, a 3 kg$\cdot$year  exposure, and  $N_{\rm int} = 180$. Left panels assume no background, while right panels include the  exponential background.
    The horizontal dashed lines mark 1.0, 0.5, and 0.1 to guide the eye. 
}
    \label{fig:b2-over-MSE}
\end{figure*}

Important indicators of the accuracy, precision and overall quality of an unfolding method are its bias  $b_j = \left<\delta\Phi_j\right>- \delta\Phi^{\rm Best\,fit}_j$, dispersion $\sigma_j= \sqrt{\left<[ \delta\Phi_j - \left<\delta\Phi_j\right>]^2\right>}$ and mean squared error MSE$_j= b_j^2 + \sigma_j^2$ (see, e.g., Ref.~\cite{Cowan:1998}).  They are shown in Figs.~\ref{fig:Relative-b-sigma-sqrtMSE}, \ref{fig:b2-variance-MSE} and \ref{fig:b2-over-MSE} for the 1~eV (upper panels) and 5 eV (lower panels) thresholds, a 3 kg$\cdot$year exposure, $N_{\rm int} = 180$, and  either no background (left panels) or exponential background (right panels), following the procedure in Appendix~\ref{AppD:MSE}.

The dispersion is about  $10$\% to 20\% of the best fit value and the bias is smaller than the dispersion at most energies, except at the low and high energy ends, indicating that the quality of the fit decreases at both ends.
The reason why the bias and dispersion are large at low energies is similar to why the upper edge of the confidence bands extend far from the best fit. The bias and dispersion are also large at the high-energy end, perhaps because truncating some of the response functions $\mc R_i(E_\nu)$ 
 in Fig.~\ref{fig:IntegratedCurlyR} at $E_\nu =2$~MeV  introduces a boundary distortion in our fit.
 
The sawtooth structure in the bias at intermediate energies in   Fig.~\ref{fig:Relative-b-sigma-sqrtMSE} is due to the piecewise constant nature of the best-fit function. On the other hand, $\left<\delta\Phi_j\right>$ is a smooth decreasing function, which is higher than the best fit in the low energy part of a constant segment of the best fit (so the bias is positive) and lower than the best fit in the higher energy part of a constant segment (so the bias is negative). This pattern repeats for each constant segment of the best fit. A different choice of data bins, with bin sizes that vary with energy may affect the bias. Also, a larger number $d$ of data bins, would likely result in a best-fit function with a larger number of downward steps (since their maximum number is $d-1$), and consequently shorter constant segments, rendering the bias closer to zero at the ends of each of these segments. 

We show
$\sigma_j^2$, $b_j^2$ and MSE$_j$ in Fig.~\ref{fig:b2-variance-MSE}, and $b_j^2/$MSE$_j$ in Fig.~\ref{fig:b2-over-MSE}. Since $b_j^2/$MSE$_j <0.5$ for most bins, the MSE is dominated by the variance. Note that we have not attempted to optimize the bias, e.g., by changing the experimental data binning or implementing variable width intervals in our ansatz.

\section{Summary}
\label{Sec:Discussion}

We developed an unfolding method by writing the function we want to determine as a finite Dirac sum, with at most $d-1$ terms, where $d$ is the number of data points.  This form is dictated by results in convex geometry~\cite{Gelmini:2017aqe}. The function to unfold is the (detector-independent) differential reactor-neutrino spectrum in the convolution that defines the CE$\nu$NS rate. The resulting integrated neutrino spectrum as a function of the lowest energy in each bin is thus piecewise constant with at most $d-1$ downward steps. The location and height of each step of the integrated flux, a total of at most 2$(d-1)$ parameters, need to be determined by likelihood maximization. A general procedure  for constructing  pointwise confidence bands was defined in Ref.~\cite{Gelmini:2017aqe}.
Although any likelihood can be maximized with a function of this kind, the best-fit function may not be unique, leading to the existence of a degeneracy band~\cite{Gelmini:2017aqe}. 

The particular extremization procedure to find  the parameters is not unique. To deal with binned data, we choose to minimize the Neyman $\chi^2$, which is equivalent to maximizing a Gaussian likelihood. This is a good approximation to the actual Poisson distribution, since the number of events in each bin is large.
Moreover, we choose to find the location and height of each step by  dividing the neutrino energy range of interest in a very large number $N_{\rm int}$ of intervals, in each of which the integrated flux takes a unique value, to be found by minimization subject to the conditions that the values be nonnegative and nonincreasing with energy (monotonicity condition). For simplicity, we choose the intervals to be equally spaced, but they can be chosen with variable width (see Appendix~\ref{AppD:Method-1}). The location of the steps is found by increasing $N_{\rm int}$ to the point at which increasing it further does not improve the goodness of the fit or the shape of the best-fit flux.

Our unfolding approach offers several distinct advantages over traditional Tikhonov regularization frameworks. First, it eliminates an important user-dependent bias. In the Tikhonov approach, selecting both the specific functional form of the regularization function and its associated regularization parameter relies heavily on empirical, subjective choices. 
Tikhonov regularization inherently introduces a systematic bias by forcing the unfolded spectrum toward a smooth function. Although the bias can be quantified, it is not explicitly included in the calculation of uncertainties. Furthermore,  the widths of the regularized bands are highly sensitive to the amount of smoothing chosen, i.e., changing the value of the regularization parameter can arbitrarily expand or contract the uncertainties. Additionally, Tikhonov regularization does not enforce physical boundaries, namely the flux can take nonphysical negative values. Also, backgrounds of complicated shape can be difficult to handle with Tikhonov regularization.  If the background has a simple shape,  a standard first-derivative penalty (which forces flat slopes) or second-derivative penalty (which forces linear gradients) works efficiently. However, if the background is complex (such as the 
exponential-plus-constant model we used) and the underlying physical gradients in the neutrino flux change rapidly -- as is characteristic of the sub-1.8 MeV differential neutrino flux probed via coherent scattering -- the Tikhonov process may require a highly complex, customized penalty matrix that accounts for the background's local curvature, or risk introducing severe, unphysical shape distortion. 

Replacing this subjective balancing act with our constrained minimization procedure removes reliance on human-tuned parameters, preserves physical boundaries of the flux, and handles backgrounds of any shape without any added complexity. The bias in our procedure is largely fixed, although it can be optimized by changing the experimental binning, a procedure also available in Tikhonov regularization. The variance and bias in our method become large at both ends of the energy interval, indicating a deterioration in the quality of the reconstruction at low and high energies.

A distinction of our method is that it explicitly reveals whether the likelihood function possesses a degenerate band of functions that maximize it. This degeneracy is an intrinsic property of the likelihood itself, which naturally broadens the confidence bands. This underlying structure is typically obscured within a standard Tikhonov regularization procedure.

A limitation of our current approach is that it fits and defines confidence bands for the integrated flux.  Further work is necessary to translate these results into a best-fit differential flux and its associated confidence bands.
Another feature is  that the resulting pointwise confidence bands exhibit a noticeable insensitivity to the background level. Through direct  testing, where we isolated the background scale while keeping all other variables constant, we verified that this behavior is an inherent consequence of our procedure. A rigorous mathematical characterization of the underlying cause of 
this issue is left for future work. 
In a Tikhonov-regularized unfolding framework, the sensitivity of the confidence bands to the background level decreases as the regularization parameter is increased. Unless excessive smoothing is enforced to flatten the spectrum, the resulting confidence bands should remain sensitive to the background scale; however,  this sensitivity was not examined in Ref.~\cite{Liao:2023kyy}.

In summary, the FDS method, based on writing the differential flux as a finite Dirac sum,  offers a conservative and objective tool to unfold the low-energy reactor neutrino flux from $\CEnNS$ data. However, we demonstrated that it not possible to  establish the existence of the neutron-capture component with currently planned experiments. The method yields results that are  comparable to traditional Tikhonov regularization frameworks, yet it completely eliminates the user-dependent arbitrariness inherent to those methods. Further investigation is necessary to develop this relatively new unfolding technique, 
originally explored in the context of direct dark matter detection, to fully characterize its capabilities and underlying statistical properties.

\acknowledgments

The authors thank R.~Cousins for helpful discussions. D.M. thanks CERN for its hospitality during the completion of this work. G.B.G. was partially supported by the US Department of Energy under Award Number DE-SC0009937. D.M. is supported in part by the U.S. Department of Energy under Grant No.~de-sc0010504. K.Y. was supported by the LANL-SoCal Hub Graduate Fellowship program and the UC SoCal Hub.

\section*{Appendices}
\appendix
\section{$\CEnNS$ cross section}
\label{AppA:cross-section}

 Natural Ge consists of $^{74}$Ge (36.5\%), $^{72}$Ge (27.5\%), $^{70}$Ge (20.6\%), $^{73}$Ge (7.7\%), and $^{76}$Ge (7.7\%)~\cite{Meija:2016isotopic}. For simplicity, and to compare our results with those of Ref.~\cite{Liao:2023kyy}, we model the target as pure $^{72}$Ge, whose mass number $A = 72$ provides a good approximation to the standard atomic weight of natural Ge, 72.63~\cite{IUPAC}. The number of neutrons in $^{72}$Ge is $N_n = A - Z = 40$ since the number of protons is  $Z = 32$.
The Standard Model $\CEnNS$ cross section is~\cite{1974PhRvD...9.1389F}
\al{
    {\p \sigma_{\rm N}\ov \p E_{R}}\fn{E_{R}, E_\nu} = {G_\text{F}^2}{{M_\text{N}}\ov 4\pi} q_\text{W}^2 \pn{1 - {{M_\text{N}}E_{R}\ov 2 E_\nu^2}}F^2\fn{q}\,,\label{crosssection}
}
where $q$ is the momentum transfer and the nuclear recoil energy is $E_R = q^2/\pn{2M_\text{N}}$. For $^{72}$Ge we take $M_{\rm N}=66.99~\GeV/c^2 \simeq 67.0~\GeV/c^2$~\cite{AME2020}, corresponding to
$1/M_{\rm N} = N_A/A_r \simeq 8.37 \times 10^{24}~\kg^{-1}$ nuclei per unit detector mass, where $N_A \simeq 6.022 \times 10^{23}~\mathrm{mol}^{-1}$ is the Avogadro number~\cite{ParticleDataGroup:2024cfk} and $A_r =71.922~\g/\mathrm{mol} \simeq 72~\g/\mathrm{mol}$ is the relative atomic mass of $^{72}$Ge~\cite{IUPAC}. Here, $G_\text{F} = 1.166\times 10^{-5}~\GeV^{-2}$ is the Fermi coupling constant~\cite{ParticleDataGroup:2024cfk}, and $q_\text{W} = N_n - \pn{1 - 4\sin^2\fn{\theta_\text{W}}}Z \simeq 38.6$ is the weak charge, which depends on the weak mixing angle $\sin^2\theta_{\tx{W}} \simeq 0.23873$~\cite{ParticleDataGroup:2024cfk}.
$F\fn{q}$ is the dimensionless nuclear form factor as a function of the momentum transfer $q$. We adopt the Klein--Nystrand parametrization~\cite{PhysRevLett.84.2330} of $F\fn{q}$, which approximates the nucleus as a uniform hard sphere distribution convolved with a Yukawa surface profile,
\al{
    F\fn{q}
    &= {3\ov \pn{qR_\text{A}}^3}
    {{\sin\fn{qR_\text{A}} - {qR_\text{A}}\cos\fn{qR_\text{A}}}\ov 1 + \pn{qa}^2}\,,
}
where $a = 0.7~\fm$ is the range of the Yukawa potential~\cite{PhysRevLett.84.2330} and $R_\text{A} = 1.2\,A^{1/3} \simeq 4.99~\fm$ is the nuclear radius~\cite{Blatt:1952ije}. Because of the low momentum transfer in $\CEnNS$ of reactor neutrinos, the theoretical rate is insensitive to the specific choice of the commonly used form factors and its uncertainties~\cite{AristizabalSierra:2019zmy}.

\section{Reactor antineutrino flux}
\label{AppB:theoretical-nu-flux}

NUCLEUS will be placed $r_1=72$~m and $r_2=102$~m from the two 4.25~GW reactor cores at the CHOOZ nuclear power plant. 
We model the differential antineutrino flux at the detector as in Ref.~\cite{Liao:2023kyy}:
\al{
    {\phi\fn{E_\nu}} &= {P\ov \tilde \epsilon} {1\ov 4\pi r_\tx{eff}^2}{\df N_\nu\ov \df E_\nu},
    \label{phi-nu-dNdE}
}
where $P \simeq 2.65\times 10^{22}~\MeV/\sec$ is the thermal power of each reactor core~\cite{Apollonio:2002gd}, and $\tilde \epsilon = 205.3~\MeV$ is the average energy released per fission~\cite{Kopeikin:2004cn,Ma:2012bm}. The effective distance of the detector to the source $r_\tx{eff}\equiv 1/\sqrt{\pn{1 / r_1^2} + \pn{1 / r_2^2}}$.

Two methods are used to calculate the reactor neutrino spectrum. The ``conversion method''~\cite{Mueller:2011nm,Huber:2011wv,Kopeikin:2021ugh} fits measured beta-electron spectra from target fissions with ``virtual branches'' (a set of about 30 hypothetical beta branches). The ``summation method'' builds spectra over a wide energy range by aggregating fission yields and decay data for isotopes from databases, but has larger uncertainties than the conversion method due to missing nuclear information; for a review see Ref.~\cite{Hayes:2016qnu}.

Contributions to ${\df N_\nu/ \df E_\nu}$ arise from the beta-decay of four main fissile isotopes 
${^{235}\tx{U}}, {^{238}\tx{U}}, {^{239}\tx{Pu}}, {^{241}\tx{Pu}}$, and from the two $\beta$ decays in the neutron-capture on ${^{238}\tx{U}}$: $^{238}$U + n $\to$ $^{239}$U $\to$ $^{239}$Np $\to$ $^{239}$Pu.
Because 0.61 neutrons are captured per fission, $1.22$ $\beta$-decay antineutrinos are produced per fission. 
The NC component, which has $E_\nu \lesssim 1.3$~MeV, is derived from the beta spectra of $^{239}$U and $^{239}$Np~\cite{Kopeikin:1997ve,Wong:2006nx}.
To compute the fission contribution to ${\df N_\nu/ \df E_\nu}$ we apply the results of the summation method below 2~MeV and the conversion method above~\cite{Liao:2023kyy}. 
Above $2~\MeV$, the spectral shapes of the $^{235}$U and $^{238}$U fission contributions are taken from Ref.~\cite{Kopeikin:2021ugh},
and for $^{239}$Pu and $^{241}$Pu from Ref.~\cite{Huber:2011wv}.
Below $2~\MeV$ we use the results of Ref.~\cite{PhysRevD.39.3378}. 
All together, the typical neutrino yields per fission of $^{235}$U, $^{238}$U, $^{239}$Pu, $^{241}$Pu and the NC process are 3.4, 0.5, 1.8, 0.4 and 1.2, respectively~\cite{Wong:2006nx}.
Because the fission and NC components have different origins and we are interested in discovering the NC component, in Fig.~\ref{fig-Int-Flux}, we show the neutrino spectra with (solid line) and without (dashed line) the latter contribution. 

\begin{figure}
    \centering
    \includegraphics[width=0.7\linewidth]{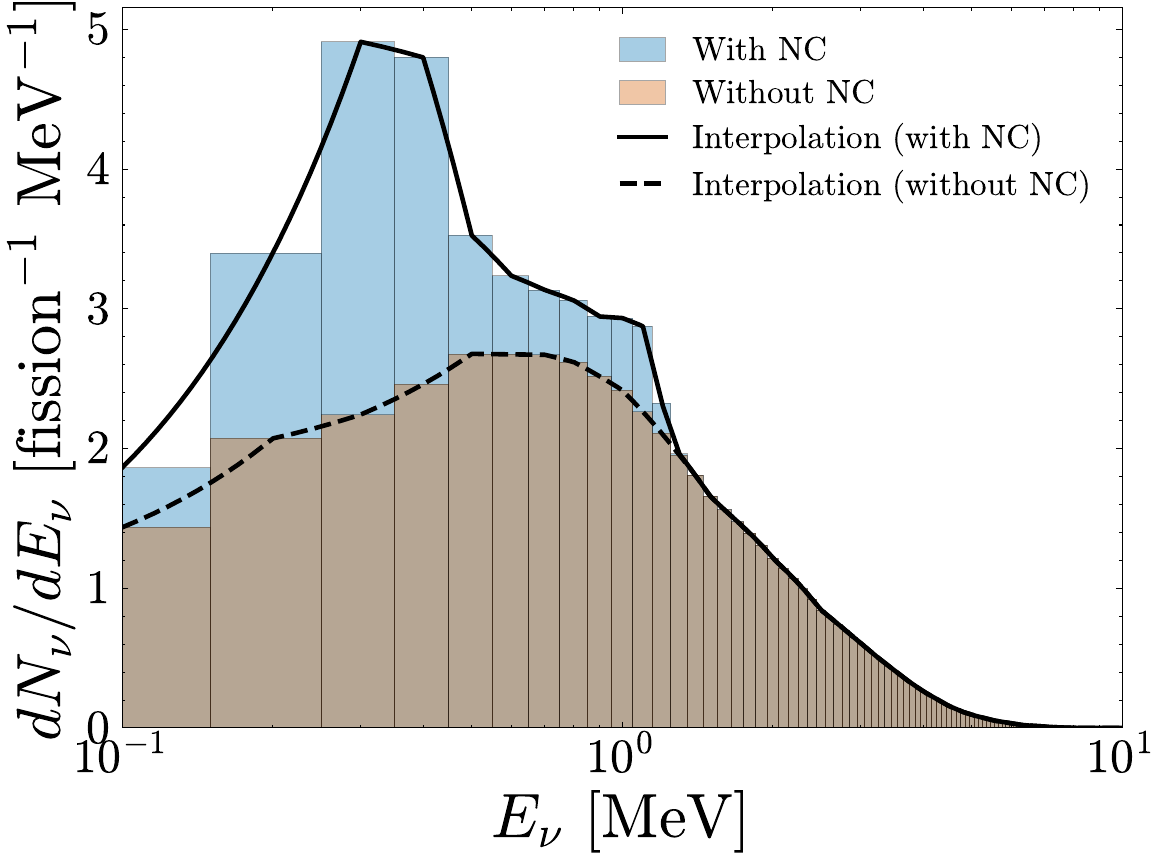}
    \caption{Predicted reactor antineutrino spectrum per fission including (blue histogram and solid curve) and excluding (orange histogram and dashed curve) the contribution of neutron-capture on $^{238}$U. Up to a constant factor defined in Eq.~\eqref{phi-nu-dNdE}, the curves show the two theoretical predictions for the differential antineutrino spectrum $\phi$.}
    \label{dNdE_histogram_interpolation}
\end{figure}

We discretize the full spectrum into energy bins with uniform width. This binned flux is then used as an input to generate $\CEnNS$ recoil spectra. 
For the rate calculation, we use a continuous spectrum obtained by piecewise-linear interpolation of the tabulated points.
Figure~\ref{dNdE_histogram_interpolation} shows the tabulated data as a histogram together with the interpolating curve, for the two cases with and without the contribution of NC on $^{238}$U.
The latter dominates the spectrum below $E_\nu \sim 1$~MeV, while they coincide above $\sim 1.3$~MeV.

\section{Alternative method}
\label{AppD:Method-1}

Although our primary goal is to reconstruct the integrated neutrino spectrum for $E_\nu <1.8$~MeV as detailed in the main text, it is possible to fit  the spectrum over the entire $E_\nu$ range where the bin-integrated response functions $\mc R_i(E_\nu)$ in Eq.~\eqref{Bin-IntegratedCurlyR}  and Fig.~\ref{fig:IntegratedCurlyR} are significantly nonzero.  While the procedure closely follows the approach used in  dark matter direct detection applications, it introduces a distinct disadvantage in our context: it forces a fit of the spectrum above 1.8~MeV, which is already well-constrained by other measurements. To implement this method, we choose an  ansatz equivalent to Eq.~\eqref{StepForm} but for $\Phi (E_\nu)$ in the wider energy ranges  $\sqbr{0.18~\MeV, 7~\MeV}$ for $E^\pr_{\rm thr} = 1~\eV$, and $\sqbr{0.41~\MeV, 7~\MeV}$ for $E^\pr_{\rm thr} = 5~\eV$:
\al{
\Phi\fn{E_\nu} = \sum_{j=1}^{N_{\rm int}} \Phi_j\,\sqbr{\Theta\fn{E_\nu - E_\nu^j} - \Theta\fn{E_\nu - E_\nu^{j+1}}}\,.
\label{StepForm-Method-1}
}
For our numerical analysis, we take either equal-width intervals or variable-width intervals with smaller widths at higher  $E_\nu$, where the $\mc R_i(E_\nu)$ are larger because they correspond to wider data bins. We find that maintaining the maximum response matrix elements $\mc R_{ij}$ corresponding to different $\mc R_i(E_\nu)$ not too different from each other improves the stability of our minimization procedure. Choosing larger interval widths at higher $E_\nu$ instead, leads to numerical problems. 

\begin{figure}
     \centering
     \includegraphics[width=0.7\linewidth]{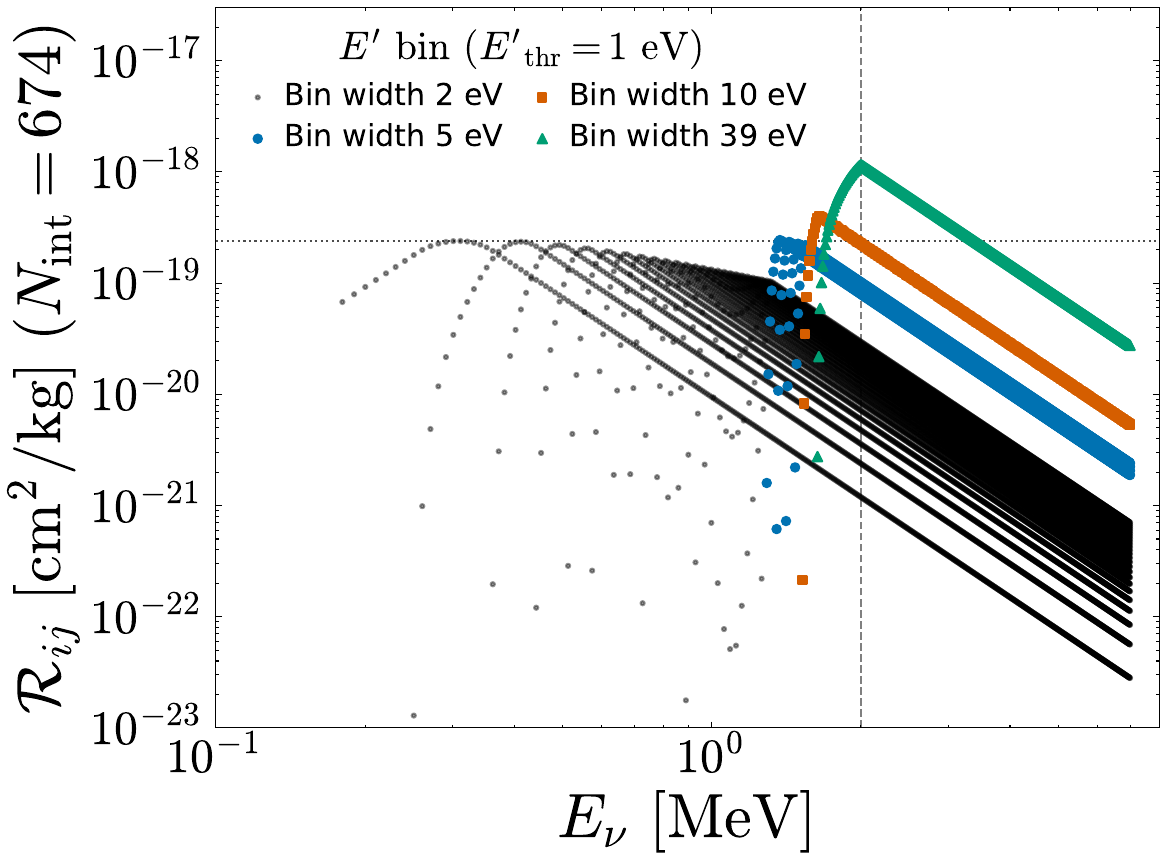}
     \caption{Response matrix elements $\mc R_{ij}$ for  $N_{\rm int}=674$ equal-width $E_\nu$ intervals within $\sqbr{0.18~\MeV, 7~\MeV}$ corresponding to a 1~eV threshold ($i= 1, \ldots , 31$ numbers the data bins, indicated by different symbols and colors, and $j= 1, \ldots, 674 $ numbers the intervals in our ansatz). The horizontal and vertical dashed lines indicate the maximum $\mc R_{1j}$ element and  2~MeV, respectively. Some colored lines appear to be continuous simply because the symbols are too close to each other. }
     \label{method1_Rij_scatter_1eV}
 \end{figure}

\begin{figure}
    \centering
    \includegraphics[width=0.45\linewidth]{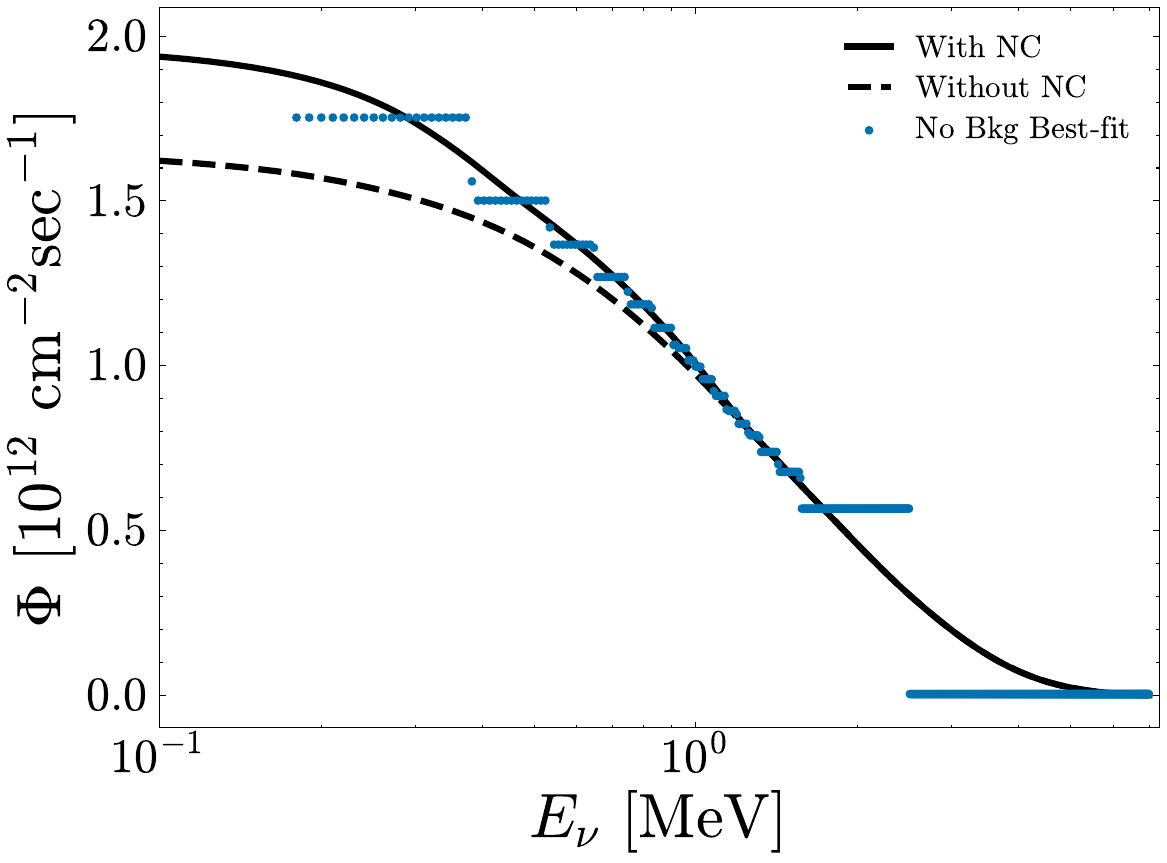}
    \includegraphics[width=0.45\linewidth]{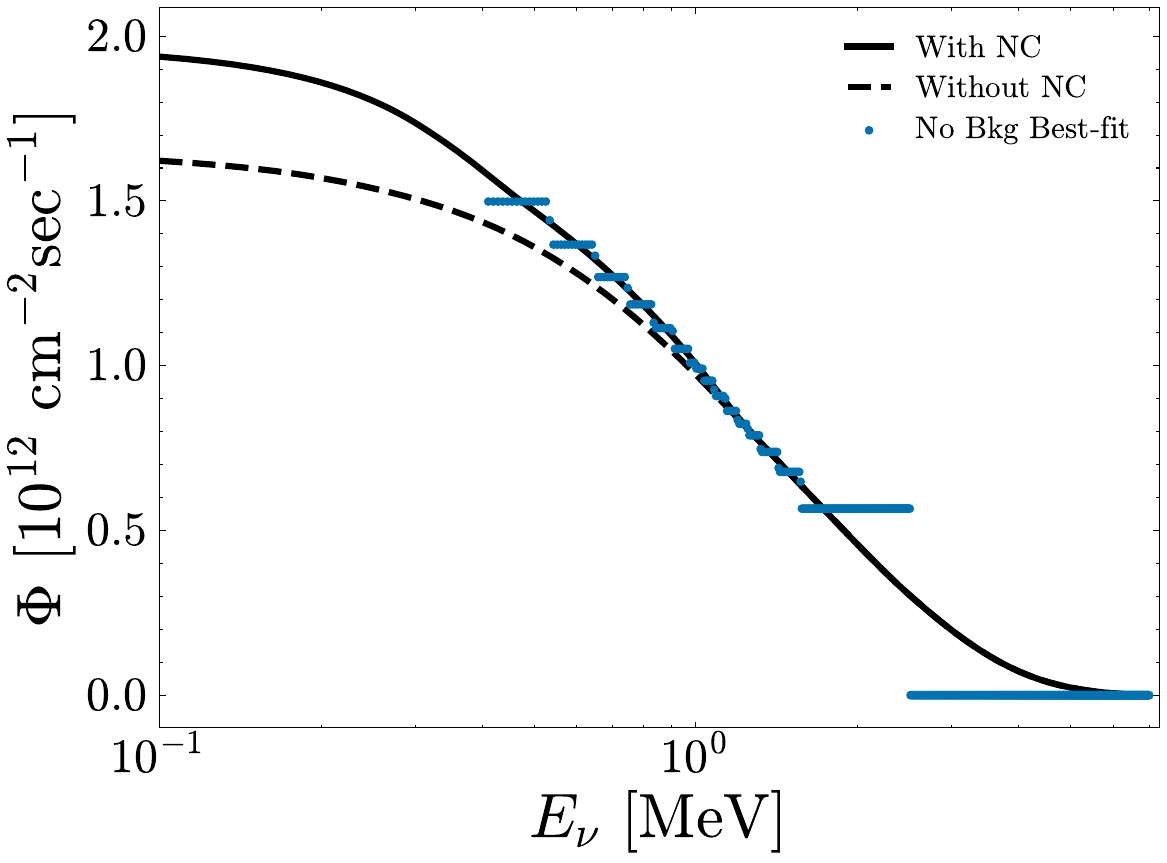}
    \caption{Piecewise-constant best-fit integrated reactor antineutrino flux $\Phi$ (blue dots) up to 7~MeV, assuming a $3~\kg\cdot$year exposure, no background,  and $E^\pr_{\rm thr} = 1~\eV$ (left panel) or $5~\eV$ (right panel) threshold, with $N_\tx{int} = 674$ and $746$, respectively.  
    The best-fit functions have 30 (left panel) and 28 (right panel) downward steps, coinciding with the maximum number $d - 1$. The  $\mc R_{ij}$ elements used for the $E^\pr_{\rm thr} = 1~\eV$ fit  are shown in Fig.~\ref{method1_Rij_scatter_1eV}.
    The black solid and dashed curves are the theoretical fluxes in Fig.~\ref{fig-Int-Flux}. 
    }
    \label{method1_bestfits}
\end{figure}

 \begin{figure}
     \centering
     \includegraphics[width=0.7\linewidth]{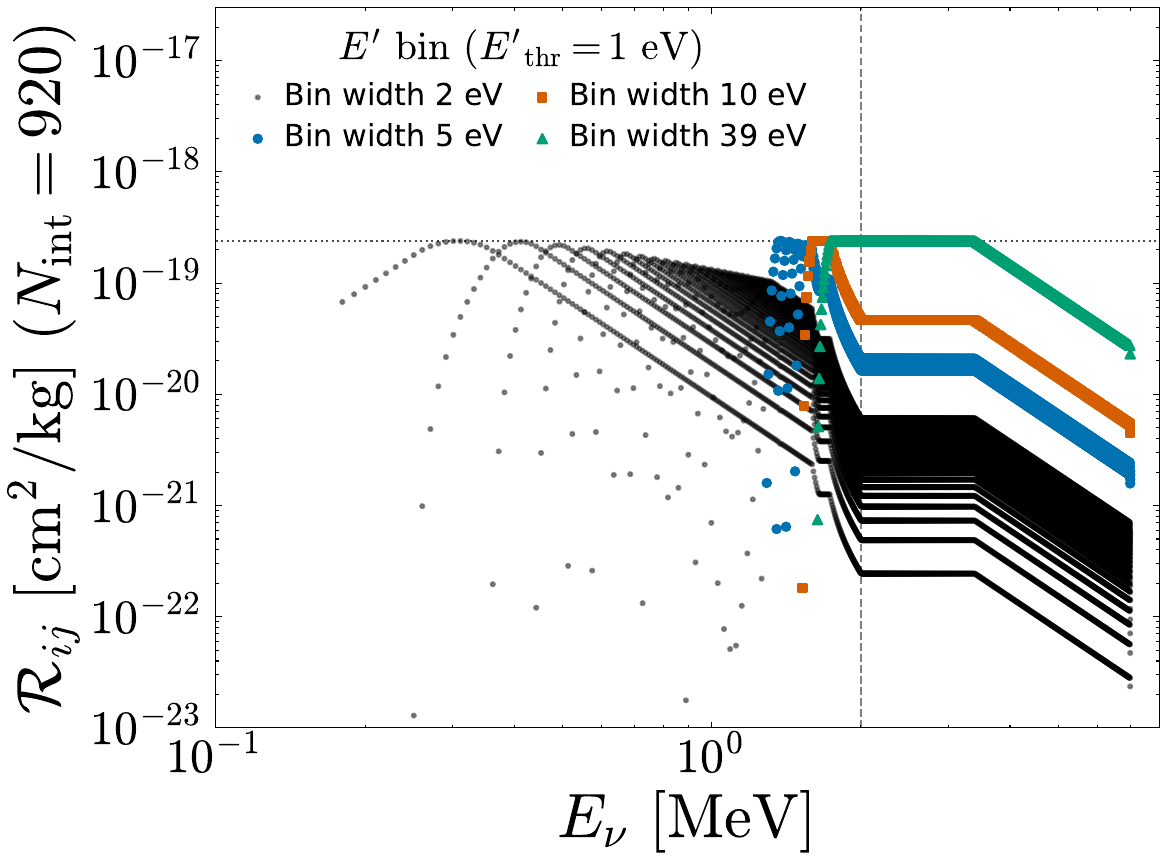}
     \caption{Similar to Fig.~\ref{method1_Rij_scatter_1eV}, with 
     $N_{\rm int}=920$ variable-width $E_\nu$ intervals within $\sqbr{0.18~\MeV, 7~\MeV}$.
     }
     \label{method1eq_Rij_scatter_1eV}
 \end{figure}


The $\{\Phi_j\}$ parameters are determined by minimizing the Neyman $\chi^2$ statistic,
\al{
    \chi^2 = \sum_i {1\over \mc E\mc O_i} \sqbr{\mc E\pn{\displaystyle\sum_{j = 1}^{N_\tx{int}}\mathcal R_{ij}\Phi_j + B_i} - \mc E\mc O_i}^2\,,
\label{chiSq-Method-1}
}
subject to the nonnegativity and nonincreasing (monotonicity) constraints,
\al{
    \Phi_j \geq 0\,, \qquad \Phi_{j+1} \leq \Phi_j\,, \qquad j = 1, \ldots, N_{\rm int} - 1\,.
\label{eq:constraints-Method-1}
}
which replace Eq.~\eqref{chiSq} and Eq.~\eqref{eq:constraints}, respectively.


Note that since we are now fitting the neutrino spectrum for $E_\nu > 2$~MeV, which includes the high-energy contributions $h_i$, these are excluded from Eq.~\eqref{chiSq-Method-1}, in contrast with Eq.~\eqref{chiSq}.

The response matrix elements obtained with equal-width intervals for a 1~eV threshold are shown in Fig.~\ref{method1_Rij_scatter_1eV}.
The resulting best-fit functions are displayed in Fig.~\ref{method1_bestfits} for a 1~eV (left panel) and 5~eV (right panel) threshold, assuming a 3~kg$\cdot$year exposure, no background,  and $N_\text{int}= 674$ and 746 uniform intervals, respectively (180 intervals below 2~MeV as in the main text, extended with the same width up to 7~MeV with 494 and 566 additional intervals above 2~MeV).\footnote{The interval width is larger for 180 equal-width intervals between [0.18~MeV, 2~MeV] for a 1~eV threshold, than for 180 equal-width intervals in the smaller energy interval [0.41~MeV, 2~MeV] for a 5~eV threshold.}

We also performed fits with variable-width intervals, chosen to be narrower if a response-matrix element computed with equal-width intervals  exceeds the largest element of the lowest energy data bin $i=1$. The response functions $\mc R_i\fn{E_\nu}$ of the wider data bins at high $E'$  are larger and peak at higher $E_\nu$, so our choice of interval widths between $\simeq 1.4~\MeV$ and 7~MeV are only about 20\% of the low $E_\nu$ width.   To make the largest element  $\mc R_{ij}$ for each $i$ above 1.4~MeV coincide by construction with the largest $i=1$ element, we have  $N_\tx{int} = 920$  and 1070 for a 1~eV and a 5~eV threshold, respectively.  The resulting $\mc R_{ij}$ for a 1~eV threshold are shown in Fig.~\ref{method1eq_Rij_scatter_1eV}. 
The best-fit functions obtained with this procedure are identical to those in Fig.~\ref{method1_bestfits}, except with a higher density of blue points.

The method detailed in the main text is better suited to our primary goal of determining the spectrum below $E_\nu = 1.8$~MeV. We do not pursue this alternative procedure further.

\section{Confidence bands}
\label{AppC:Monte-Carlo}

 We find the mapping from a $\Delta\chi^2_{\min}$ value to a confidence level by a Monte Carlo method,  using the parametric bootstrap. For an account of this method see Ref.~\cite{Demortier:2007zz} and Section~13 of Ref.~\cite{Cousins:2018tiz}, 
  and references therein. Here we describe the construction of confidence bands at $95.45\%$ CL, corresponding to 2$\sigma$. Other CL values follow the same pattern. 

\begin{itemize}
\item{Step 1: We perform a  global fit to the dataset $\{\mathcal{O}_i\}$, $i = 1, \ldots, d$, following the procedure introduced in Section~\ref{Sec:CEvNS-rate}, in which the $\chi^2$ of Eq.~\eqref{chiSq} is minimized assuming the ansatz for $\delta\Phi$ in Eq.~\eqref{StepForm}, subject only to the positivity and monotonicity conditions, $\delta\Phi_j \geq 0$ and $\delta\Phi_{j+1} \leq \delta\Phi_i$ in Eq.~\eqref{eq:constraints}. We refer to this as ``free'' minimization (as opposed to the ``constrained'' minimization in Step~2). This yields the best-fit piecewise constant function, with values $\{{\delta \hat\Phi}_j\}=\{\delta\Phi^{\rm Best\,fit}_j\}$ for all the  intervals  $j= 1, \ldots, N_{\rm int}$, and the global minimum $\chi^2$, denoted $\chi^2_{\min}$. 
}

\item{Step 2: To probe whether a particular fixed value $\delta\Phi_j^*\neq \delta\Phi^{\rm Best\,fit}_j$ in the $j^{\rm th}$ interval lies within the $2\sigma$ band, we first perform a minimization while keeping $\delta\Phi_j$ pinned to $\delta\Phi_j^*$ (we call this a ``constrained minimization''). This produces ${\delta\hat{\hat \Phi}}_k$ values for all $k \neq j$ (which are our nuisance parameters). The minimum $\chi^2$ value with $\delta\Phi_j^*$ fixed is necessarily larger than its global minimum value $\chi^2_{\min}$, with a difference \mbox{$\Delta\chi^2_{\min}(\delta\Phi_j^*)\ge 0$.}}

\item{Step 3: With the $\delta\Phi$ obtained in Step~2, namely $\delta\Phi_j^*$ and $ {\delta\hat{\hat \Phi}}_k$ for $k \neq j$,  we compute the total predicted  number of mock events in each experimental bin, $\{N_i^{\rm exp}(\delta\Phi_j^*)\}$ for $i= 1, \dots , d$, which also includes  the $\CEnNS$ contribution from the high-energy neutrinos, $h_i$, and the background $B_i$.}

This predicted dataset  defines the parent distribution from which we generate $N_{\rm boot}$ pseudodatasets, or ``bootstrap iterations'',  $\{N_i\}_a$, $a = 1, \dots, N_{\rm boot}$, where each $N_i$ is drawn independently from a Gaussian centered on $N_i^{\rm exp}(\delta\Phi_j^*)$ with variance $N_i^{\rm exp}(\delta\Phi_j^*)$, and independently $h_i$ ($B_i$) are drawn from Gaussian distributions centered at $h_i$ ($B_i$) with variance $h_i$ ($B_i$).
The only exception is the comparison with Ref.~\cite{Liao:2023kyy} in Fig.~\ref{Flat Bkg 5eV 3kgyrs}. There we only fluctuate the total number of events ${N_i}$ because Ref.~\cite{Liao:2023kyy} does not include independent fluctuations of $h_i$ and $B_i$.

We set $N_{\rm boot} = 500$, for which our determination of the 2$\sigma$ bands has a 1\% error. Each pseudo-experiment has a chance to fall within the desired confidence level, described by a binomial distribution with the ``success probability'' $p=X/N$, where $X$ is the number of successes and $N$ is the total number of trials. For large $N$, and $p$ not very close to 0 or 1, $p$ is Gaussian distributed.
The standard deviation of $p$ is $\sigma_p=\sqrt{p(1-p)/N}$, and in our case  $N= N_{\rm boot}$. To achieve a certain $\sigma_p$, the number of trials needed is $N=p(1-p)/\sigma_p^2$. With $N_{\rm boot}=500$ we have $\sigma_p=0.021$ for $p=0.68$ (1$\sigma$) and $\sigma_p=0.010$ for $p=0.95$ (2$\sigma$).

\item{Step 4: For each pseudodataset we perform two independent fits, always with the conditions in Eq.~\eqref{eq:constraints},  a free minimization  and a constrained minimization which holds $\delta\Phi_j$ fixed at $\delta\Phi_j^*$. The latter must have a larger minimum $\chi^2$ than the former, with a difference $\Delta\chi^2_{\min a}(\delta\Phi_j^*)\ge 0$, where the index $a$ labels the pseudodata sets, $a= 1, \ldots, N_{\rm boot}$.
We sort the values of $\Delta\chi^2_{\min a}(\delta\Phi_j^*)$ in increasing order and find 
$\Delta\chi^2_{\rm cut}(\delta\Phi_j^*)$, defined as the value that has  
95.45\% of them below it (for the 2$\sigma$ limit). 
In other words, the empirical distribution of the $N_{\rm boot}$ $\Delta\chi^2_{\min a}(\delta\Phi_j^*)$ values is the Monte Carlo realization of this test statistic under the hypothesis $\delta\Phi_j = \delta\Phi_j^*$, and reading off its $95.45^{\rm th}$ percentile gives the cutoff $\Delta\chi^2_{\rm cut}(\delta\Phi_j^*)$. }

\item{Step 5: We now compare this cutoff against the value $\Delta\chi^2_{\min}(\delta\Phi_j^*)$
obtained from the original data in Step~2. Then,  $\delta\Phi_j^*$ lies inside the $95.45\%$ pointwise confidence band at $E_\nu^j$ iff $\Delta\chi^2_{\min}(\delta\Phi_j^*) \leq \Delta\chi^2_{\rm cut}(\delta\Phi_j^*)$.}

 \item{Step 6: Scanning the values of $\delta\Phi_j^*$ on a fine grid then delineates the upper and lower edges of the band at $E_\nu^j$, and repeating the construction for each $j$ produces the full pointwise confidence band. In practice, we compute the band at several energies and interpolate the boundaries in between.}

\end{itemize}

\begin{figure*}[t]
    \centering
    \includegraphics[width=\linewidth]{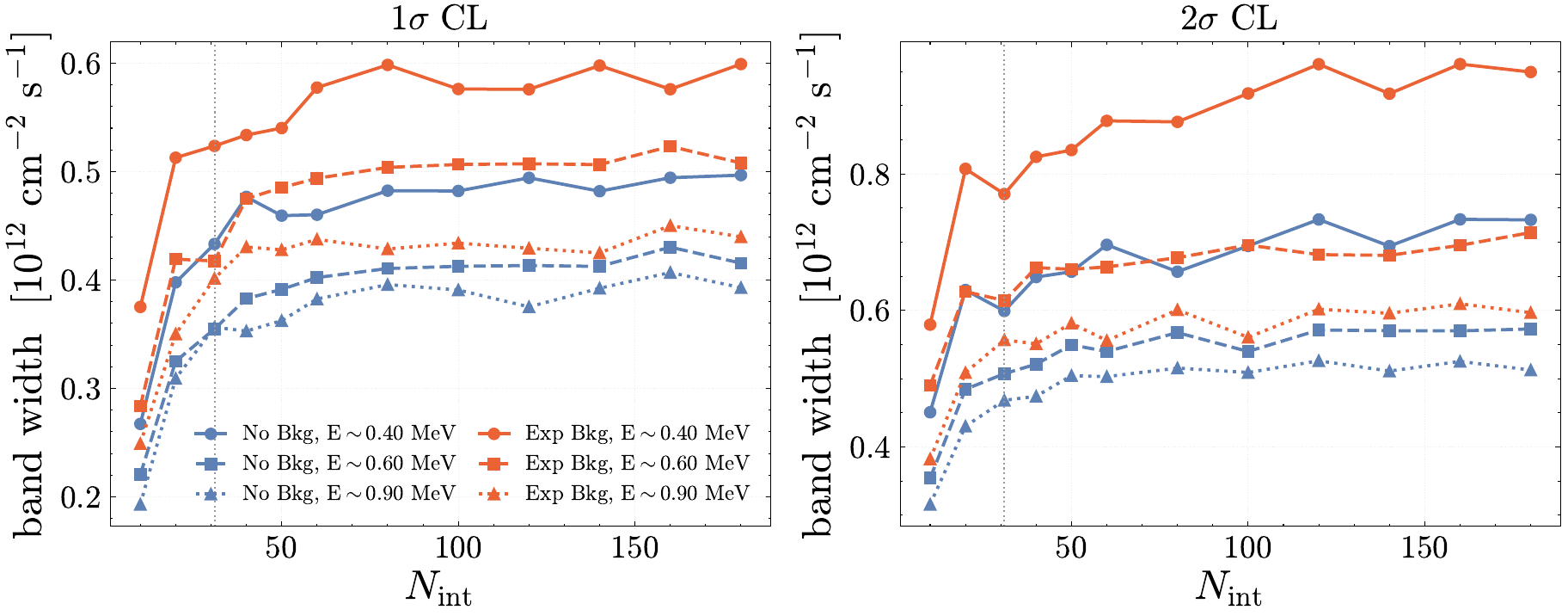}
    \caption{Confidence band width as a function of $N_{\rm int}$ for the 1$\sigma$ (left) and 2$\sigma$ (right) bands and three different energies, without a background (blue) and with the exponential background (red). The vertical dotted line marks the number $d$ of data points. Only for values of $N_{\rm int}$ much larger than $d$ do the widths plateau.
    }
    \label{width_vs_n}
\end{figure*}

The $2\sigma$ pointwise bands shown in Figs.~\ref{OnlyHighEnergyNeutrino}, \ref{1sigma2sigma}, \ref{1eV 50kgyrs}, \ref{5eV 330kgyrs}, and \ref{Flat Bkg 5eV 3kgyrs} are constructed by this procedure with the $95.45^{\rm th}$ percentile. The $1\sigma$ band in  Fig.~\ref{1sigma2sigma} corresponds to the $68.27^{\rm th}$ percentile.

The degeneracy band is obtained by identifying the values of $\delta\Phi_j^*$ for which $\Delta\chi^2_{\min}(\delta\Phi_j^*) \leq 10^{-3}$. The arbitrary limit of $10^{-3}$ is chosen to be much smaller than any significant confidence level value. The degeneracy band does not change significantly if other small upper limits are chosen.

Another issue worth clarifying is the choice of the number of intervals $N_{\rm int}$ in our ansatz. Due to the many minimizations carried out, we tried to use the smallest value of $N_{\rm int}$ possible, while obtaining stable bands. The dependence of the width of the 1$\sigma$ (left panel) and 2$\sigma$ (right panel) bands on the number of intervals, $N_{\rm int}$,  is shown in Fig.~\ref{width_vs_n}. In all cases, the width of the band increases significantly for $N_{\rm int}\lesssim d=31$ (the number of data bins $d$ is indicated by the vertical dotted line) and then plateaus, becoming essentially independent of $N_{\rm int}$ above about 2$d$. We set  $N_{\rm int} =80$. This means that the band width becomes stable and subdividing the $E_\nu$ domain into finer intervals does not produce further improvement above this value. As expected, the 2$\sigma$ band widths are systematically larger than the corresponding 1$\sigma$ widths. Including the expected background (red curves) broadens the bands relative to the background-free case (blue curves) under otherwise identical conditions, while the band width decreases with increasing energy.

\section{Bias, dispersion and mean squared error}
\label{AppD:MSE}

We describe the calculation of the bias, variance, and mean squared error (MSE) for each energy interval $j = 1, \ldots, N_{\rm int}$.
We start with the global fit to our original data  $\{\mathcal{O}_i\}$, $i= 1, \ldots, d$,  to obtain the best fit $ {\delta \hat\Phi}_j = \delta\Phi^{\rm Best\,fit}_j$ -- where $\delta\Phi^{\rm Best\,fit}_j$ is the name we used earlier and is shown in Figs.~\eqref{OnlyHighEnergyNeutrino} to~\eqref{Flat Bkg 5eV 3kgyrs} for $ N_{\rm int}=180$ --
following Step~1 of Appendix~\ref{AppC:Monte-Carlo}. The set $\{\delta\hat{\Phi}_j\}= \{\delta\Phi^{\rm Best\,fit}_j\}$ represents the unknown ``true'' neutrino flux model.

With this model we generate $N_{\rm boot}=10^4$ pseudodatasets, with the same binning as our original dataset, and find the best fit for each of them, 
$\delta{\Phi}^u_j$ for $u = 1, \ldots, N_{\rm boot}$.
For each $j$ we compute the bias as the difference between the average $\left<{\delta\Phi}_j\right>$ of 
$\delta{\Phi}^u_j$, and the ``true'' value $\delta\Phi^{\rm Best\,fit}_j$,
\begin{eqnarray}
    b_j \equiv \left<{\delta\Phi}_j\right>-\delta \hat{\Phi}_j = \left[\frac{1}{N_{\rm boot}}\sum_u\delta\Phi_j^u\right]- \delta\Phi^{\rm Best\,fit}_j\,,
\end{eqnarray}
 the variance of the $N_{\rm boot}$ values, 
\begin{eqnarray}
    \hat{\sigma}^2_j \equiv \frac{1}{(N_{\rm boot}-1)}\sum_u\left[\delta\Phi_j^u-\left<{\delta\Phi}_j\right> \right]^2\,,
\end{eqnarray}
and the MSE,
\begin{eqnarray}
    {\rm MSE}_j\equiv b_j^2+\sigma_j^2\,.
\end{eqnarray}

\bibliographystyle{JHEP}
\bibliography{ref.bib}

\providecommand{\href}[2]{#2}\begingroup\raggedright\begin{thebibliography}{10}

\bibitem{Li:2001ha}
H.-B. Li and H.~T. Wong, {\it {Sensitivities of low-energy reactor neutrino
  experiments}},  {\em J. Phys. G} {\bf 28} (2002) 1453--1468,
  [\href{https://arxiv.org/abs/hep-ex/0111002}{{\tt hep-ex/0111002}}].

\bibitem{COHERENT:2017ipa}
{\bf COHERENT} Collaboration, D.~Akimov {\em et~al.}, {\it {Observation of
  Coherent Elastic Neutrino-Nucleus Scattering}},  {\em Science} {\bf 357}
  (2017), no.~6356 1123--1126, [\href{https://arxiv.org/abs/1708.01294}{{\tt
  1708.01294}}].

\bibitem{Colaresi:2022obx}
J.~Colaresi, J.~I. Collar, T.~W. Hossbach, C.~M. Lewis, and K.~M. Yocum, {\it
  {Measurement of Coherent Elastic Neutrino-Nucleus Scattering from Reactor
  Antineutrinos}},  {\em Phys. Rev. Lett.} {\bf 129} (2022), no.~21 211802,
  [\href{https://arxiv.org/abs/2202.09672}{{\tt 2202.09672}}].

\bibitem{NUCLEUS:2017htt}
{\bf NUCLEUS} Collaboration, R.~Strauss {\em et~al.}, {\it {The $\nu$-cleus
  experiment: A gram-scale fiducial-volume cryogenic detector for the first
  detection of coherent neutrino-nucleus scattering}},  {\em Eur. Phys. J. C}
  {\bf 77} (2017) 506, [\href{https://arxiv.org/abs/1704.04320}{{\tt
  1704.04320}}].

\bibitem{NUCLEUS:2019igx}
{\bf NUCLEUS} Collaboration, G.~Angloher {\em et~al.}, {\it {Exploring $\hbox
  {CE}\nu \hbox {NS}$ with NUCLEUS at the Chooz nuclear power plant}},  {\em
  Eur. Phys. J. C} {\bf 79} (2019), no.~12 1018,
  [\href{https://arxiv.org/abs/1905.10258}{{\tt 1905.10258}}].

\bibitem{Liao:2023kyy}
J.~Liao, H.~Liu, and D.~Marfatia, {\it {How to measure the reactor neutrino
  flux below the inverse beta decay threshold with CE\ensuremath{\nu}NS}},
  {\em Phys. Rev. D} {\bf 108} (2023), no.~3 033002,
  [\href{https://arxiv.org/abs/2302.10460}{{\tt 2302.10460}}].

\bibitem{NUCLEUS:2026pnv}
{\bf NUCLEUS} Collaboration, H.~Abele {\em et~al.}, {\it {Prospect of the
  NUCLEUS experiment at Chooz for coherent elastic neutrino-nucleus scattering
  and new physics searches}},  {\em Phys. Rev. D} {\bf 114} (2026), no.~1
  012016, [\href{https://arxiv.org/abs/2603.24450}{{\tt 2603.24450}}].

\bibitem{Gelmini:2017aqe}
G.~B. Gelmini, J.-H. Huh, and S.~J. Witte, {\it {Unified Halo-Independent
  Formalism From Convex Hulls for Direct Dark Matter Searches}},  {\em JCAP}
  {\bf 12} (2017) 039, [\href{https://arxiv.org/abs/1707.07019}{{\tt
  1707.07019}}].

\bibitem{Fox:2010bz}
P.~J. Fox, J.~Liu, and N.~Weiner, {\it {Integrating Out Astrophysical
  Uncertainties}},  {\em Phys. Rev. D} {\bf 83} (2011) 103514,
  [\href{https://arxiv.org/abs/1011.1915}{{\tt 1011.1915}}].

\bibitem{Fox:2011qd}
P.~J. Fox, J.~Liu, D.~Tucker-Smith, and N.~Weiner, {\it {An Effective Z'}},
  {\em Phys. Rev. D} {\bf 84} (2011) 115006,
  [\href{https://arxiv.org/abs/1104.4127}{{\tt 1104.4127}}].

\bibitem{Frandsen:2011gi}
M.~T. Frandsen, F.~Kahlhoefer, C.~McCabe, S.~Sarkar, and K.~Schmidt-Hoberg,
  {\it {Resolving astrophysical uncertainties in dark matter direct
  detection}},  {\em JCAP} {\bf 01} (2012) 024,
  [\href{https://arxiv.org/abs/1111.0292}{{\tt 1111.0292}}].

\bibitem{Gondolo:2012rs}
P.~Gondolo and G.~B. Gelmini, {\it {Halo independent comparison of direct dark
  matter detection data}},  {\em JCAP} {\bf 12} (2012) 015,
  [\href{https://arxiv.org/abs/1202.6359}{{\tt 1202.6359}}].

\bibitem{DelNobile:2013cta}
E.~Del~Nobile, G.~B. Gelmini, P.~Gondolo, and J.-H. Huh, {\it {Halo-independent
  analysis of direct detection data for light WIMPs}},  {\em JCAP} {\bf 10}
  (2013) 026, [\href{https://arxiv.org/abs/1304.6183}{{\tt 1304.6183}}].

\bibitem{DelNobile:2013cva}
E.~Del~Nobile, G.~Gelmini, P.~Gondolo, and J.-H. Huh, {\it {Generalized Halo
  Independent Comparison of Direct Dark Matter Detection Data}},  {\em JCAP}
  {\bf 10} (2013) 048, [\href{https://arxiv.org/abs/1306.5273}{{\tt
  1306.5273}}].

\bibitem{Fox:2014kua}
P.~J. Fox, Y.~Kahn, and M.~McCullough, {\it {Taking Halo-Independent Dark
  Matter Methods Out of the Bin}},  {\em JCAP} {\bf 10} (2014) 076,
  [\href{https://arxiv.org/abs/1403.6830}{{\tt 1403.6830}}].

\bibitem{Gelmini:2015voa}
G.~B. Gelmini, A.~Georgescu, P.~Gondolo, and J.-H. Huh, {\it {Extended Maximum
  Likelihood Halo-independent Analysis of Dark Matter Direct Detection Data}},
  {\em JCAP} {\bf 11} (2015) 038, [\href{https://arxiv.org/abs/1507.03902}{{\tt
  1507.03902}}].

\bibitem{Feldstein:2014gza}
B.~Feldstein and F.~Kahlhoefer, {\it {A new halo-independent approach to dark
  matter direct detection analysis}},  {\em JCAP} {\bf 08} (2014) 065,
  [\href{https://arxiv.org/abs/1403.4606}{{\tt 1403.4606}}].

\bibitem{Georgescu:CoddsDM}
A.~Georgescu, {\it {\it CoddsDM: Comparing data from direct searches for Dark
  Matter}}, .

\bibitem{IUPAC}
T.~Prohaska {\em et~al.}, {\it Standard atomic weights of the elements 2021
  (iupac technical report)},  {\em Pure and Applied Chemistry} {\bf 94} (2022),
  no.~5 573--600.

\bibitem{NUCLEUS:2024sensitivity}
{\bf NUCLEUS} Collaboration, R.~Wagner {\em et~al.}, {\it {Sensitivity
  enhancement techniques for cryogenic calorimeters in the NUCLEUS
  experiment}},  \href{https://arxiv.org/abs/2603.28276}{{\tt 2603.28276}}.

\bibitem{Ricochet:2024CryoCube}
{\bf Ricochet} Collaboration, C.~Augier {\em et~al.}, {\it {First demonstration
  of 30 eVee ionization energy resolution with Ricochet germanium cryogenic
  bolometers}},  {\em Eur. Phys. J. C} {\bf 84} (2024), no.~2 212,
  [\href{https://arxiv.org/abs/2306.00166}{{\tt 2306.00166}}].

\bibitem{Demortier:2007zz}
L.~Demortier, {\it {P values and nuisance parameters}},  in {\em {PHYSTAT-LHC
  Workshop on Statistical Issues for LHC Physics}}, pp.~23--33, 2007.

\bibitem{Cousins:2018tiz}
R.~D. Cousins, {\it {Lectures on Statistics in Theory: Prelude to Statistics in
  Practice}},  \href{https://arxiv.org/abs/1807.05996}{{\tt 1807.05996}}.

\bibitem{Cowan:1998}
G.~Cowan, {\em Statistical Data Analysis}.
\newblock Clarendon Press, Oxford, 1998.

\bibitem{Meija:2016isotopic}
J.~Meija, T.~B. Coplen, M.~Berglund, W.~A. Brand, P.~De~Bi{\`e}vre,
  M.~Gr{\"o}ning, N.~E. Holden, J.~Irrgeher, R.~D. Loss, T.~Walczyk, and
  T.~Prohaska, {\it Isotopic compositions of the elements 2013 (iupac technical
  report)},  {\em Pure and Applied Chemistry} {\bf 88} (2016), no.~3 293--306.

\bibitem{1974PhRvD...9.1389F}
D.~Z. {Freedman}, {\it {Coherent effects of a weak neutral current}},  {\em
  PRD} {\bf 9} (Mar., 1974) 1389--1392.

\bibitem{AME2020}
M.~Wang, W.~J. Huang, F.~G. Kondev, G.~Audi, and S.~Naimi, {\it {The AME 2020
  atomic mass evaluation (II). Tables, graphs and references}},  {\em Chinese
  Phys. C} {\bf 45} (2021), no.~3 030003.

\bibitem{ParticleDataGroup:2024cfk}
{\bf Particle Data Group} Collaboration, S.~Navas {\em et~al.}, {\it {Review of
  particle physics}},  {\em Phys. Rev. D} {\bf 110} (2024), no.~3 030001.

\bibitem{PhysRevLett.84.2330}
S.~R. Klein and J.~Nystrand, {\it Interference in exclusive vector meson
  production in heavy-ion collisions},  {\em Phys. Rev. Lett.} {\bf 84} (Mar,
  2000) 2330--2333.

\bibitem{Blatt:1952ije}
J.~M. Blatt and V.~F. Weisskopf, {\em {Theoretical nuclear physics}}.
\newblock Springer, New York, 1952.

\bibitem{AristizabalSierra:2019zmy}
D.~Aristizabal~Sierra, J.~Liao, and D.~Marfatia, {\it {Impact of form factor
  uncertainties on interpretations of coherent elastic neutrino-nucleus
  scattering data}},  {\em JHEP} {\bf 06} (2019) 141,
  [\href{https://arxiv.org/abs/1902.07398}{{\tt 1902.07398}}].

\bibitem{Apollonio:2002gd}
{\bf CHOOZ} Collaboration, M.~Apollonio {\em et~al.}, {\it {Initial results
  from the CHOOZ long baseline reactor neutrino oscillation experiment}},  {\em
  Eur. Phys. J. C} {\bf 27} (2003) 331--374,
  [\href{https://arxiv.org/abs/hep-ex/0301017}{{\tt hep-ex/0301017}}].

\bibitem{Kopeikin:2004cn}
V.~I. Kopeikin, L.~A. Mikaelyan, and V.~V. Sinev, {\it {Reactor as a source of
  antineutrinos: Thermal fission energy}},  {\em Phys. Atom. Nucl.} {\bf 67}
  (2004) 1892--1899.

\bibitem{Ma:2012bm}
X.~B. Ma, W.~L. Zhong, L.~J. Wang, Y.~X. Chen, and J.~Cao, {\it {Improved
  calculation of the energy release in neutron-induced fission}},  {\em Phys.
  Rev. C} {\bf 88} (2013) 014605, [\href{https://arxiv.org/abs/1212.6625}{{\tt
  1212.6625}}].

\bibitem{Mueller:2011nm}
T.~A. Mueller {\em et~al.}, {\it {Improved Predictions of Reactor Antineutrino
  Spectra}},  {\em Phys. Rev. C} {\bf 83} (2011) 054615,
  [\href{https://arxiv.org/abs/1101.2663}{{\tt 1101.2663}}].

\bibitem{Huber:2011wv}
P.~Huber, {\it {On the determination of anti-neutrino spectra from nuclear
  reactors}},  {\em Phys. Rev. C} {\bf 84} (2011) 024617,
  [\href{https://arxiv.org/abs/1106.0687}{{\tt 1106.0687}}]. [Erratum:
  Phys.Rev.C 85, 029901 (2012)].

\bibitem{Kopeikin:2021ugh}
V.~Kopeikin, M.~Skorokhvatov, and O.~Titov, {\it Reevaluating reactor
  antineutrino spectra with new measurements of the ratio between u235 and
  pu239 \ensuremath{\beta} spectra},  {\em Phys. Rev. D} {\bf 104} (2021),
  no.~7 L071301, [\href{https://arxiv.org/abs/2103.01684}{{\tt 2103.01684}}].

\bibitem{Hayes:2016qnu}
A.~C. Hayes and P.~Vogel, {\it {Reactor Neutrino Spectra}},  {\em Ann. Rev.
  Nucl. Part. Sci.} {\bf 66} (2016) 219--244,
  [\href{https://arxiv.org/abs/1605.02047}{{\tt 1605.02047}}].

\bibitem{Kopeikin:1997ve}
V.~I. Kopeikin, L.~A. Mikaelyan, and V.~V. Sinev, {\it {Spectrum of electronic
  reactor anti-neutrinos}},  {\em Phys. Atom. Nucl.} {\bf 60} (1997) 172--176.

\bibitem{Wong:2006nx}
{\bf TEXONO} Collaboration, H.~T. Wong {\em et~al.}, {\it {Search of neutrino
  magnetic moments with a high-purity germanium detector at the Kuo-Sheng
  nuclear power station}},  {\em Phys. Rev. D} {\bf 75} (2007) 012001,
  [\href{https://arxiv.org/abs/hep-ex/0605006}{{\tt hep-ex/0605006}}].

\bibitem{PhysRevD.39.3378}
P.~Vogel and J.~Engel, {\it Neutrino electromagnetic form factors},  {\em Phys.
  Rev. D} {\bf 39} (Jun, 1989) 3378--3383.

\end{thebibliography}\endgroup


\end{document}